\documentclass[useAMS,usenatbib,twocolumn]{mnras}
\usepackage{times}
\usepackage{graphics,epsfig}
\usepackage{graphicx,amsmath,amsopn,multicol}
\usepackage{amssymb}
\usepackage{txfonts}
\usepackage{psfrag}
\usepackage{setspace}
\usepackage{amsfonts}                                                                              
\usepackage{multirow}
\expandafter\def\csname ver@subfig.sty\endcsname{}
\usepackage{placeins}
\usepackage{subfig}
\usepackage{mathrsfs}
\usepackage[utf8]{inputenc}
\usepackage{setspace}
\usepackage{mathtools}
\usepackage{grffile}
\usepackage{amsmath}
\usepackage{longtable}
\usepackage{graphics}
\usepackage{verbatim}
\usepackage[usenames]{color}
\usepackage{float}
\usepackage{bm}
\usepackage{epstopdf}

\newcommand{\U}{\mathbfit{U}}
\newcommand{\Ug}{\mathbfit{U}_{\rm g}}
\newcommand{\Ui}{\mathbfit{U}_{\rm i}}
\newcommand{\Un}{\mathbfit{U}^{'}}
\newcommand{\W}{\textit{W} (\bm{\theta})}
\newcommand{\A}{\textit{A} (\bm{\theta})}
\newcommand{\R}{\textit{R} (\bm{\theta})}

\newcommand{\tv}{\bm{\theta}}
\newcommand{\V}{\mathcal{V}}
\newcommand{\sig}{\mathcal{S}}
\newcommand{\N}{\mathcal{N}}
\newcommand{\T}{\delta \textit{T} (\bm{\theta})}
\newcommand{\Ts}{\delta \textit{T}_{s} (\bm{\theta})}

\title[SNR auto \& cross power spectrum]{The  auto and cross angular power spectrum of the Cas A supernova remnant in  radio and X-ray}
\author[P. Saha et al.]{Preetha Saha,$^{1}$\thanks{preetha@phy.iitkgp.ernet.in} Somnath Bharadwaj,$^{1}$\thanks{somnath@phy.iitkgp.ac.in} Susmita Chakravorty, Nirupam Roy,$^{2}$ \newauthor Samir Choudhuri,$^{3}$ Hans Moritz Günther,$^{4}$ Randall K. Smith$^{5}$  \\~\\ $^{1}$ Department of Physics and Centre for Theoretical Studies, Indian Institute of Technology, Kharagpur 721302, India\\
$^{2}$ Department of Physics, Indian Institute of Science, Bangalore 560012, India\\
$^{3}$ Astronomy Unit, Queen Mary University of London, Mile End Road, London E1 4NS, United Kingdom\\
$^{4}$ MIT, Kavli Institute for Astrophysics and Space Research, 77 Massachusetts Avenue, Cambridge, MA 02139, USA\\
$^{5}$ Harvard-Smithsonian Center for Astrophysics, 60 Garden St, Cambridge, MA 02138, USA
}

\begin{document}
\date{Accepted yyyy month dd. Received yyyy month dd; in original form yyyy month dd}

\pagerange{\pageref{firstpage}--\pageref{lastpage}} 
\pubyear{2021}

\maketitle

\label{firstpage}

\begin{abstract}
The  shell type  supernova remnant (SNR) Cas A exhibits structures at  nearly all  angular scales.  Previous studies show the angular power spectrum $(C_{\ell})$ of the  radio emission to be a broken power law, consistent with MHD turbulence. 
The break has been identified with the transition from 2D to 3D turbulence at the angular scale corresponding to the shell thickness.  Alternatively, this can also be explained as  
2D inverse cascade driven by  energy injection from knot-shock interactions.
Here we present $C_{\ell}$ measured from archival VLA $5$GHz (C band) data, and Chandra X-ray data in the energy ranges ${\rm A}=0.6-1.0 \, \,  {\rm  keV}$ and  ${\rm B} =4.2-6.0 \, \, {\rm keV}$, both of which are continuum dominated. The different emissions all trace fluctuations in the underlying plasma and possibly also the  magnetic field, and we expect them to be correlated. 
We quantify this using
the cross $C_{\ell}$ between the different emissions. We find that X-ray B is strongly correlated with both radio and X-ray A, however X-ray A is only very weakly correlated with radio. This supports a picture where X-ray A is predominantly thermal  bremsstrahlung whereas X-ray B is a composite of thermal  bremsstrahlung and non-thermal synchrotron emission.  The various $C_{\ell}$  measured here, 
all show a broken power law behaviour. However, the slopes are typically shallower than those in radio and the position of the break also corresponds to smaller angular scales.
 These findings provide observational inputs regarding the nature of turbulence and the emission mechanisms in Cas A. 
\end{abstract}


\section{Introduction}
A supernova explosion is characterized by the release of a vast amount of kinetic energy from a point in space. This energy in the form of a shock wave travels ahead of the ejected stellar material from the explosion. Bounded by the shock wave expanding through the surrounding interstellar medium (ISM), a supernova remnant (SNR) consists of the stellar ejecta and the interstellar material swept along its way. The temporal evolution of the remnant is described in terms of four evolutionary stages \citep{Woltjer1972}. 
In the first phase the mass swept up by the shock is much less than the mass of the stellar ejecta and  the stellar ejecta expands at the initial shock speed. This stage is referred to as the  `ejecta-driven' (ED) phase. Since the energy radiated is still very small compared to the kinetic energy released during the explosion, the remnant undergoes adiabatic expansion  and enters its next `Sedov-Taylor' (ST) phase. In this phase, the mass swept up by the shock begins to dominate the ejecta.
Based on a simple spherically symmetric model, the self-similarity solutions \citep{sedov1946,Taylor1949} can explain only the adiabatic phase of the evolution.
However, the one-dimensional self-similarity does not remain valid for the later phases of the adiabatic evolution due to the interaction of the ejecta with the inhomogeneities in the ISM \citep{Chevalier1977}.
The evolution ceases to be adiabatic once  the radiative loss become significant. We now have the `pressure driven snowplow' (PDS) phase where the radiative shock is driven by the interior pressure. Here the blast wave decelerates and the dense shell surrounding the hot interior cools down. In the fourth phase,  the interior gas starts losing energy by `snowplowing' or moving the shell through the ISM. This is referred to as the `momentum driven snowplow' (MDS) phase. In an uniform density circumstellar medium (CSM), the temporal  evolution  of the forward shock radius follows a power-law $t^m$, with values  $m=1,\frac{2}{5},\frac{2}{7}, \frac{1}{4}$ for these four phases respectively \citep{Cioffi1988}. For Cas A SNR, it is found that CSM has a density varying inversely with the square of the shock radius \citep{HwangLaming09}. This indicates that the remnant is in its ST phase of evolution as the swept up mass is much greater than that of the ejecta, but most of the X-ray emission comes from the ejecta.

Besides the expansion studies, SNRs are studied over the entire electromagnetic spectrum from radio to high energy gamma bands. Each frequency band of observation offers a different physical insight to the understanding of supernovae and SNRs. Considering any particular frequency, every SNR exhibits a variety of overall morphology as well as rich structures over a wide range of scales. SNRs can be used to probe the global properties of the galaxy as well as the local environment in different regions. Unlike the SNRs which are unresolved at extragalactic distances, SNRs in our Galaxy provide us an opportunity to investigate their structures in finer details \citep{Green2019}. 

 Although there have been many multi wavelength observations of the Galactic SNRs, there have been very few systematic studies to quantify the statistical properties of the fine-scale angular structures associated with the remnant. \cite{roy09}  have  estimated the angular power spectrum of the observed intensity fluctuations for two Galactic SNRs, Cas A and Crab using the visibility data from Very Large Array (VLA) at frequencies $~1.5$ GHz (L band) and $~5$ GHz (C band). The resultant angular power spectra of both the SNRs show a power law with  index $-3.24\pm 0.03$. 
  However, for the Cas A SNR a break in the power law was observed at large angular scales  where  the index was found to change from  $-3.2$ to $-2.2$. This break  was attributed  to a transition from 3D to 2D turbulence on scales larger than the shell thickness of Cas A. A similar transition was  not observed  for the Crab SNR which has a  filled centre structure. In \cite{roy09}, the power spectrum was estimated using pairwise  correlations of the measured visibilities (the  Bare Estimator of \citealt{Choudhuri2014}). In a recent  work \citet{Saha2019} have estimated 
 the angular power spectrum of the Kepler SNR using an improved estimator the Tapered Gridded Estimator (TGE) \citep{Choudhuri2016b}. This estimator reduces the residual point source contamination from regions of the sky away from the target source by tapering the primary beam response through an appropriate convolution in the visibility domain. The angular power spectrum   was measured  in both  L and C bands using VLA  data. In both cases the power spectrum measured across the  
 angular multipole $\ell$ range $(1.9-6.9)\times10^4$ was   found 
 to be a broken power law  with a break at $\ell_{b}=3.3\times10^4$, and  
  power law index of $-2.84$ and $-4.39$ before $(\ell < \ell_{b})$  and after $(\ell > \ell_{b})$
    the break respectively.  The C and L band results were found to differ at $\ell >6.91 \times10^4$.
     In this $\ell$ range the 
  L band  power spectrum was found to flatten out at a value which is consistent with model predictions for the expected  residual point source foreground contribution. However, the  C band  power spectrum was found to follow a power law with index $-3.07$. 
  Considering the synchrotron emitting plasma of the SNR, 
  the slope of $-2.84$ found at large angular scales
  is consistent with 2D Kolmogorov turbulence, whereas at small angular scales the C band power law slope $-3.07$ is roughly consistent with 3D Kolmogorov turbulence. The intermediate $\ell$ range with a  steep $-4.39$ power law behavior was attributed to the complex morphology of the  SNR. Furthermore, the angular power spectrum estimates of Cas A and Crab obtained using TGE were found to be similar with the earlier results reported in \cite{roy09}. In another recent work \cite{shimoda2018} have analyzed VLA L band image of the Tycho SNR to estimate the two-point correlation function of the specific intensity fluctuations. They have reported a Kolmogorov-like magnetic energy spectrum showing a scaling $r^{2/3}$. 
  \cite{Pavan20} have carried out a similar analysis for Cas A  using $410-460$ MHz  
  Giant Metrewave Radio Telescope (GMRT) data, their results are consistent with the above mentioned  Tycho SNR results.

In this paper we revisit Cas A  considering not only radio but also X-ray observations  of this SNR. 
For the present analysis we have analysed archival Chandra X-ray data (Chandra observation ID: $4638$).
Cassiopeia A or Cas A (G111.7$-$2.1) is located in the constellation Cassiopeia in the second Galactic quadrant.   Cas A is a relatively young Galactic SNR which has been extensively studied since its discovery. At the radio frequencies, Cas A SNR shows a clear shell like structure with compact emission knots. It is one of the strongest radio sources in the sky with a flux density of $2720\pm50\,$Jy at $1\,$GHz.  The shell structure subtends an angular diameter of $5^{'}$ at a distance of approximately $3.4\,$kpc \citep{Reed1995}. The shell thickness estimated from the radial  brightness  profile  at  radio  frequencies  is  found  to  be approximately $30^{''}$. It is most likely a remnant of a late 17th century supernova \citep{Fesen2006}. \cite{Krause1195} suggests that CasA was a type IIb supernova. The spectroscopic observations of Cas A SNR at X-ray frequencies reveal that it is a strong X-ray source with the presence of different elements in the supernova ejecta. Each of these elements produces X-rays within narrow energy ranges. In X-ray waveband, the overall emission structure is an approximately circular region of diffuse emission which surrounds an inner ring of clumpy emission \citep{Charles1977,Fabian1980,Jansen1988}. The circular region has a protrusion at the northeast.

In this paper we have extended the line of investigation started  in \citet{roy09} and  
 continued in \cite{Saha2019} to  present the angular power spectrum of Cas A SNR both at low  radio frequency and at high X-ray frequency. The radio power spectrum arises mainly from the fluctuations in the non-thermal synchrotron radiation. A convective instability due to the interaction of the ISM and the ejecta makes enough turbulent energy available to account for the observed synchrotron emission \citep{Gull1973a}.  The emissivity of the synchrotron radiation depends on $\sim\,n_{e}\,B^{\frac{p+1}{2}}$  where $n_{e}$ is  the electron number density, $B$  is  the magnetic field  and  $p$ is the power-law index of the electron spectrum. Considering the emission from the SNRs at X-ray frequency, this can be either thermal bremsstrahlung or non-thermal synchrotron or a combination of both.
 For Cas A SNR \cite{Allen97} have shown evidence for  a non thermal high energy tail in the X-ray continuum spectrum which  is likely to be produced by synchrotron radiation. 
 In addition to the synchrotron component, the high energy tail can also have contribution from non-thermal bremsstrahlung \citep{Asvarov90,Bleeker01,Grefenstette_2015}.
 Later, \cite{heldervink08} argued that about $54\%$ of the overall continuum emission ($4-6$ keV) of Cas  A is non thermal synchrotron radiation, while the rest is thermal  bremsstrahlung. While the radio power spectrum from the earlier works already sheds some light on the statistics of the small scale turbulent fluctuations in the electron density and magnetic field, we expect  the X-ray angular power spectrum to provide independent additional inputs on this. Further, we may expect the fluctuations in the radio and non thermal X-ray emission to be correlated as both are  related to  the electron density and also the magnetic field. In this paper we have also estimated the cross-correlation angular power spectrum between the fluctuations in the radio and the X-ray emission.

 Cross-correlation is a commonly used methodology in signal and image processing. The noise and other artefacts like extraneous background or foreground emission affecting the two different signals  are expected to be  uncorrelated, and their contribution can be avoided if we  cross-correlate the two signals.  Considering the radio observations used for our analysis, the contribution  from various   
 foregrounds like the extragalactic point sources and the diffuse Galactic synchrotron emission (DGSE) both increase if the observing frequency is reduced. Guided by this 
  we have chosen the VLA C band visibility data of Cas A SNR over that of the L band for the  analysis presented here. A brief outline of the paper follows. 

The details of the archival VLA radio data and Chandra  X-ray data used here 
are briefly outlined in Section \ref{sec:data}. The methodology of the power spectrum estimation and its error are described in Section \ref{sec:meth}.  We report the results in Section \ref{sec:res} and modeling of the radio power spectrum in \ref{sec:model_radio}, followed by its discussion and conclusion in Section \ref{sec:disc}.

\section{Data}
\label{sec:data}
\begin{figure*}
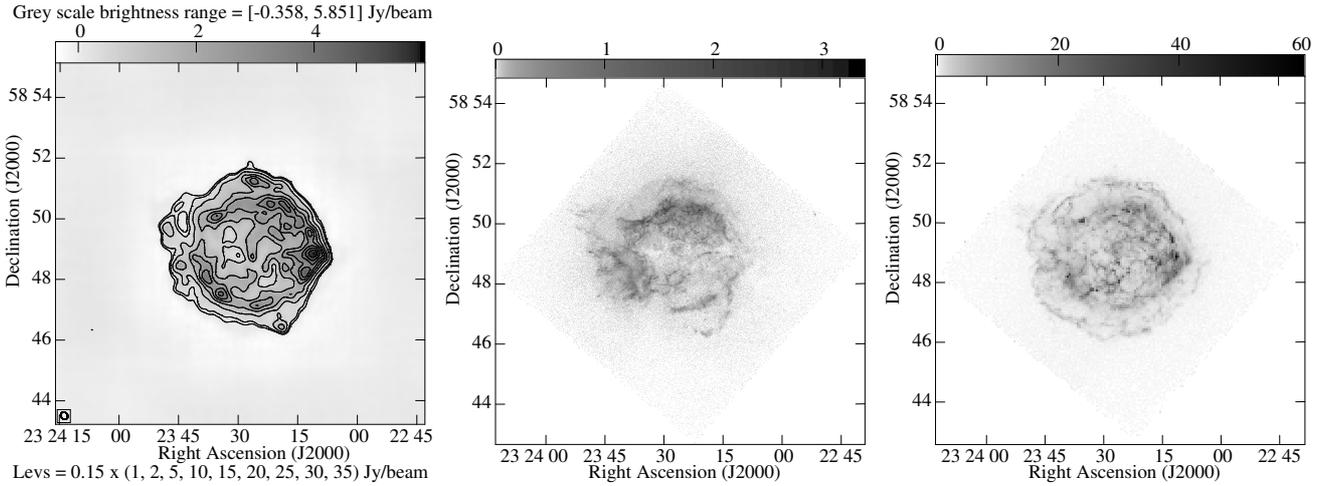


 \includegraphics[scale=0.3]{CasAD_aips_2.ps}	\includegraphics[scale=0.3]{CasAX0.6to1.ps} \includegraphics[scale=0.3]{CasAX4.2to6.ps}
 \caption{Images of Cas A SNR in Radio (left) from the VLA archival visibility data in  C band D configuration, and  X-ray from Chandra archival  data  
spanning  $0.6-1.0$ keV (X-Ray A, middle) and $4.2-6.0$ keV (X=ray B, right). The images are all in J2000 coordinates.} 
 \label{fig:AIPSCasA}

\end{figure*}

We have used the  Very Large Array (VLA) archival C band ($5\,$GHz) multi configuration \textit{uv} data (Project code: AR0435) of the Cas A supernova remnant (SNR). \citet{roy09}  have calibrated and used this data to estimate the angular power spectrum of the Cas A SNR, and the reader is referred to Table~1 of the above mentioned  paper for further details of the data. To  visualise  the source structure the left panel of Figure \ref{fig:AIPSCasA} here shows a CLEANed radio image  made from  the calibrated visibilities  using standard AIPS\footnote{NRAO Astrophysical Image Processing System, a commonly used software for radio data processing} tasks. We see that the Cas A SNR subtends approximately $5^{'}\times 5^{'}$ in the radio image. We note that  
 the  radio image here shows the specific intensity $I(\tv)$ in units of $\mathrm{Jy\,beam^{-1}}$. 
 This can be converted to   brightness temperature $T(\tv)$ through $T(\tv)=(2\,k_B/\lambda^{2})^{-1}\, I(\tv)$ in the  Rayleigh-Jeans limit which is applicable here.

\begin{figure}

 \includegraphics[width=0.52\textwidth,scale=0.3]{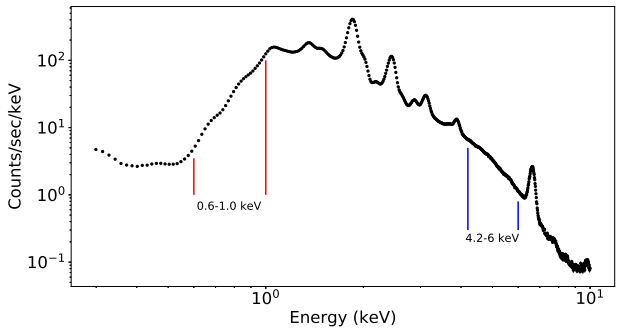}	
 \caption{Log-log plot of X-ray spectrum of Cas A SNR using Chandra archival data (observation ID: $4638$). The background has been substracted here. We have marked the lower (X-ray A, marked with red lines) and the higher (X-ray B, marked with blue lines) energy ranges that we have used to derive our images for the X-ray analysis, in this paper.}
 \label{fig:xspec} 

\end{figure}

For X-ray, we have analysed Advanced CCD Imaging Spectrometer (ACIS-S) non-grating archival Chandra data of Cas A. The data was observed on $14$th April, $2004$ as a part of observation ID $4638$ and has a long exposure time of $164.5\,$ks. The field of view (FoV) corresponds to a large region of angular extent $16.9\arcmin \times 8.3\arcmin$. We analyze the downloaded data using Chandra Interactive Analysis of Observations (CIAO) version $4.10$ \citep{ciao}. We create a reprocessed event$2$ file from the downloaded event$1$ file using \texttt{chandra\_repro}. 
Using the event$2$ file, we generate the  X-ray spectrum which is shown in Figure $\ref{fig:xspec}$. 
For the subsequent analysis we have identified  two different energy ranges  which are free of any strong emission lines, namely X-ray A and X-ray B (or simply A and B) which respectively spans $0.6-1.0$ keV and  $4.2-6.0$ keV  (Figure $\ref{fig:xspec}$).   We expect  A 
to be dominated by  thermal emission while  B is expected to be a mixture  of thermal and non-thermal emission. The middle and right panels of Figure \ref{fig:AIPSCasA} respectively show the X-ray images corresponding to A and B. 
 Here from the X-ray images, we consider $J(\tv)$ which refers to the X-ray photon flux per unit energy interval per solid angle. We note that $J(\tv)$ is analogous to the specific intensity $I(\tv)$ which refers to the energy flux per frequency interval per solid angle \citep{bradt_2008}. As seen by comparing the different images in  Figure $\ref{fig:AIPSCasA}$,   the angular size of Cas A is similar in 
 radio as well as X-ray A and  B. While the radio and X-ray B images are relatively more symmetric,  
the circular structure in X-ray A  is incomplete with the appearance of an arc on the southwestern side. The inner ring and the protrusion of the remnant in the northeast are also more prominent 
in the X-ray A image.

\section{Methodology}
\label{sec:meth}

\subsection{Radio power spectrum}
In radio interferometric observations, the measured quantity is a set of complex visibilities $\V_i$ which are sensitive to the angular fluctuations in the sky signal. In addition to the sky signal component $\sig(\Ui)$, the measured visibility $\V_i$ has a system noise $\N$ component
\begin{equation}
 \V_{i}=\sig(\Ui)+\N_{i} \,.
 \label{eq:v1}
\end{equation} 
The sky signal component $\sig(\Ui)$ is the Fourier transform of the product of the telescope's primary beam pattern $\A$ and the  brightness temperature fluctuations $\T$ on the sky 
\begin{equation}
    \label{s1}
     \sig(\Ui)=Q_{\nu} \int d^2 \tv \, e^{2\pi i \Ui \cdot\tv} \, \A \, \T 
\end{equation}
 where  $Q_{\nu}=2 k_B/\lambda^2$ is the conversion factor from brightness temperature  to specific intensity in the Rayleigh Jeans limit and $k_B$ is the Boltzmann constant. Here we assume that $\T$ is a particular realisation of a statistically homogeneous and isotropic Gaussian random field whose statistical properties are completely quantified by the angular power spectrum $C^{TT}_{\ell}$. In the flat sky approximation adopted here,  $C^{TT}_{\ell}$ which is the  angular power spectrum of brightness temperature fluctuations 
 is defined  using 
 \begin{equation}
       \label {C_l}
    \langle \Delta \tilde{T} (\U) \, \Delta \tilde{T}^{*}(\Un) \rangle  = \delta_D^{2}(\U-\Un)\,C^{TT}_{\ell=2\pi U}
 \end{equation}
 where $\Delta \tilde{T} (\U)$ is the Fourier transform of $\T$, $\delta_D^{2}(\U-\Un)$ is the 2D Dirac delta function and the angular brackets $\langle ... \rangle$ denote an ensemble average with respect to different realizations of the Gaussian random field. 
 
 The angular extent of the target source is usually smaller compared to that of $\A$.  In addition to the sky signal from the target source, the measured visibilities $ \V_{i}$ may have significant contributions from other sources which lie within the angular extent of $\A$. Therefore, it is advantageous  to restrict the sky response to a small region around the target source. The Tapered Gridded Estimator (TGE) \citep{Choudhuri2014,Choudhuri2016a} achieves this by tapering the sky response with a suitable window function $\W$. TGE restricts the sky response by convolving the measured visibilities  with $\tilde{w}(\U)$ which is  the Fourier transform of $\W$. This visibility based estimator for the radio frequency data allow us to bypass any additional complications due to imaging and deconvolution, which are quite challenging for strong extended source like Cas A SNR. TGE also grids the visibilties in order to reduce the volume of computations. The   convolved gridded $\V_{cg}$ are given by 
 \begin{equation}
 \label{V_con&grid}
\V_{cg}=\sum_{i} \tilde{w}(\Ug-\Ui)\V_{i}
\end{equation} 
where $\Ug$ refers to the baseline corresponding to the different grid points (labeled $g$)  of a  rectangular grid in the \textit{uv} plane.
The size of the rectangular $\textit{uv}$ grid is set by the choice of the baseline range $U_{max}$ and $U_{min}$. This is chosen such that the grid can incorporate the visibility data of both the largest configuration and the shortest configuration of VLA. We have chosen a grid spacing within which the convolution in eq. \eqref{V_con&grid} is well represented. The uniform grid size and spacing allows us to collapse the visibility data of the different VLA configurations into a single grid. The angular power spectrum estimator is defined as
   \begin{equation}
   \label{eqn:TGE_V_m}
   \hat{E}_{g}=(M_{g})^{-1}\times\bigg(|\V_{cg}|^{2}-\sum_{i} |\tilde{w}(\Ug-\Ui)|^{2}|\V_{i}|^{2}\bigg)
   \end{equation}
   where $M_{g}$ is a normalization constant and the second term in the {\it r.h.s.} is introduced to subtract out the noise bias arising from the noise contribution present in each visibility.   Here $ \langle\hat{E}_{g}\rangle$ gives an unbiased estimate of the angular auto power spectrum $C^{TT}_{\ell}$ at the angular multipole $\ell_{g}=2 \pi \mid \Ug \mid$. 
   
   The value of $M_g$ is calculated using simulated visibilities. We have simulated visibilities  
   corresponding to an unit angular power spectrum (UAPS, $C^{UAPS}_{\ell}=1$) and having the same baseline distribution as the observational  data. These simulations incorporate only the sky signal contribution, with no system noise. We use an ensemble of such simulations 
   to evaluate the normalization constant 
     \begin{equation}
     M_{g}=\bigg \langle \, \big (|\V_{cg}|^{2}-\sum_{i} |\tilde{w}(\Ug-\Ui)|^{2}|\V_{i}|^{2} \big )\, \bigg \rangle UAPS \,.
     \label{eq:mg}
   \end{equation}
   We apply the visibility based TGE to the radio visibility data of the SNR to estimate the angular auto power spectrum $C^{TT}_{\ell}$. 
   We have binned the estimated $C^{TT}_{\ell}$ values into $\ell$ bins of equal logarithmic interval.

   The  details of the  TGE angular power spectrum estimator are  discussed in \cite{Choudhuri2014,Choudhuri2016a}, whereas  
   the detailed methodology for the TGE based  
   $C^{TT}_{\ell}$ estimation for  several supernova remnants (Cas A, Crab and Kepler) is given in Section $3.1$ of \cite{Saha2019}.

\subsection{X-ray power spectrum}
\label{sec:x-ray}
For the  X-ray data of the Cas A SNR we consider  $J(\tv)$ which, as discussed in Section~\ref{sec:data}, is analogous to the specific intensity $I(\tv)$.  Here we use $J^A(\tv)$ and
$J^B(\tv)$ to refer to X-ray A and B respectively. Considering A, 
the X-ray image  (middle panel of Figure~\ref{fig:AIPSCasA}) presents us with values of $J^A(\tv)$ on a pixelized image of angular dimensions $L \times L=\Omega$ subtending a solid angle $\Omega$.  We calculate the Fourier components $\Delta J^A (\U)$ of the image, and calculate the X-ray auto angular power spectrum $C^{AA}_{\ell}$  using 
\begin{equation}
    C^{AA}_{\ell=2\pi U}=\Omega^{-1}\, |\Delta J^A (\U)|^{2}
\label{eq:x-ps}
\end{equation}
 The power spectrum  $C^{BB}_{\ell}$ was estimated in exactly the same way using $J^B(\tv)$.

 We have also estimated the cross-power spectrum between the two different X-ray energy bands using 
\begin{equation}
    C^{AB}_{\ell=2\pi U}=\Omega^{-1}\, \mid [\Delta J^A (\U) ] \,  [\Delta J^B (\U)]^* \mid \,.
\label{eq:xc-ps}
\end{equation} 

The estimated power spectra were all individually binned 
in $\ell$ bins of equal logarithmic interval.

\subsection{Radio X-ray Cross power spectrum}
We now discuss how we have cross-correlated the radio visibility data with the X-ray A and the X-ray B image data to estimate the cross power spectrum $C^{TA}_{\ell}$ and $C^{TB}_{\ell}$ respectively. Considering the X-ray A, we have an exposure corrected image which provides us with $\Delta J^A(\U)$. This is available at uniformly sampled $\U$ values, and we may interpret the measured $\Delta J^A(\U)$ as directly pertaining to different Fourier components of the X-ray structure of the Cas A SNR.  In contrast, the visibility data $\V_{i}$ is only available at the $\U_{i}$ values corresponding to the VLA baseline distribution.  Typically, the baseline distribution $\U_{i}$   does not span the entire Fourier domain of our interest. Further, we cannot interpret the measured visibilities directly as the Fourier components of $\T$ due to the primary beam pattern $\A$ which appears in   eq.~(\ref{s1}). Note that field of view of the Chandra X-ray image is larger than the angular extent of the  VLA primary beam pattern $\A$.
In addition, we have to also consider the window function  $\W$ if we deal with the convolved, gridded visibilities 
$\V_{cg}$ (eq.~\ref{V_con&grid}).  It is necessary to account for these aspects of the radio data when estimating the cross power spectrum. 
In order to overcome these issues, we have converted the X-ray A data into X-ray A visibilities  $\V_{{A}_{i}}$ 
using 
\begin{equation}
    \label{sx1}
     \V_{{A}_{i}}= \int d^2 \tv \, e^{2\pi i \Ui \cdot\tv} \, \A \, \delta \textit{J}^A (\bm{\theta}) .
\end{equation} 
which are exactly  analogous to the radio visibility data. Here  $\A$ is the VLA primary beam pattern, and  the X-ray A visibilities have been  computed at the $\U_i$ values corresponding to the VLA baselines distribution.  We have used the X-ray A visibilities for the cross-correlation analysis. 

We have used  the TGE to estimate the cross-angular power spectrum $C^{TA}_{\ell}$. In order to  maintain a similar sky  response from the source in both the radio and the X-ray A,  the  X-ray A visibilties $\V_{{A}_{i}}$ were convolved with the same window function  as the radio data (eq.~\ref{V_con&grid}) 
\begin{equation}
 \label{Vx_con&grid}
\V_{{A}_{cg}}=\sum_{i} \tilde{w}(\Ug-\Ui)\V_{{A}_{i}} \,,
\end{equation} 
 and the  grid also is identical to that used for the radio data.  We generalize eq.~(\ref{eqn:TGE_V_m}) to define an estimator for the 
cross power spectrum 
\begin{equation}
   \label{eqn:TGE_V_cross}
   \hat{E}_{g}=(M_{g})^{-1}  \, \mid \V_{A_{cg}} \, \V^*_{cg} \mid 
   \end{equation}
where  $M_{g}$ is the same proportionality constant as in eq. (\ref{eq:mg}). The noise in the radio and the X-ray A visiblities are uncorrelated, and it is not necessary to consider the noise bias here. We follow exactly same procedure to calculate the cross spectrum $C^{TB}_{\ell}$ using the measured radio visibilties $\V_{i}$ and the estimated X-ray B visibilities $\V_{{B}_{i}}$.
 We present the bin averaged  values of the angular cross power spectra  $C^{TA}_{\ell}$ and $C^{TB}_{\ell}$.

\subsection{Interpreting the power spectra}
To interpret the radio sky signal, the observed brightness temperature fluctuation $\T$ of the remnant can be represented as 
\begin{equation}
\T=\R [\bar{T}_{s} + \Ts]
\label{eq:snr.a}
\end{equation}
where $\bar{T}_{s}$ is the mean and $\Ts$ is the fluctuating component. $\R$ is a dimensionless profile function of the remnant which captures the angular profile of the remnant. This $\R$ modulates both $\bar{T}_{s}$ and $\Ts$ and cuts off the emission beyond the finite size of the SNR.  We assume $\Ts$ to be the outcome of a statistically homogeneous and isotropic Gaussian random process whose statistical properties are completely specified by the angular power spectrum $C_{\ell}$. A similar statistical interpretation is described in \cite{CM_V1952} in the context of turbulent ISM of our Galaxy.  

The angular power spectrum $C^{TT}_{\ell}$ estimated from $\T$ is related to $C_{\ell}$ 
(which corresponds to $\Ts$  )  through a convolution 
\begin{equation}
C^{TT}_{2 \pi \mid \U \mid} =\int d^2 \U^{'} \, \mid \tilde{r}(\U-\U^{'}) \mid^2  \, C_{2 \pi \mid \U^{'} \mid} 
\label{eq:snr.b}
\end{equation}
where $\tilde{r}(\U)$ is the Fourier transform of $\R$. Considering a power law of the form $C_{\ell} \propto A \, \ell^{\beta}$ with a negative power law index $(\beta <0)$,  we find that the estimated $C^{TT}_{\ell}$ has the same slope as $C_{\ell}$  at large $\ell$ $(\gg \ell^T_m)$. In this range, $C^{TT}_{\ell}$ and $C_{\ell}$ differ only by 
\begin{equation}
C_{2 \pi \mid \U \mid} = B^{-1}\, C^{TT}_{2 \pi \mid \U \mid} 
\label{eq:snr.c}
\end{equation}
 the proportionality constant  $B=\left[ \int d^2 \tv \, \mid \R \mid^2 \right]=\left[ \int d^2 \U^{'} \, \mid \tilde{r}(\U^{'}) \mid^2 \right]$ \citep{dutta2009}. However, at small $\ell$ $(\le \ell^T_m)$, $C^{TT}_{\ell}$ flattens out and the convolution (eq. \ref{eq:snr.b}) introduces a break at $\ell^T_m$. The value of $\ell^T_m$ is inversely proportional to the angular extent of the SNR. Depending on the shape of $\R$ and power law index $\beta$, the value of $\ell^T_m$ can vary.

 The above interpretation has been extended to the rest of the estimated power spectra $C^{AA}_{\ell}$, $C^{BB}_{\ell}$, $C^{TA}_{\ell}$, $C^{TB}_{\ell}$ and $C^{AB}_{\ell}$ as well. We have modeled the angular profile of the SNR   as a  Gaussian of the form  $\R =e^{-{\theta^{2}/\theta_{r}^{2}}}$. For each data the value of $\theta_{r}$ was chosen separately so as to correctly reproduce the observed values of $\ell_m$.

\subsection{Error estimation}
The error estimates $\delta C_{\ell}$ for $C^{TT}_{\ell}$ is relatively straight forward, and this has been discussed in \citet{Saha2019} and references therein. 
Here the sample variance which is the statistical fluctuation inherent to the sky signal as well as the system noise both contribute to the errors. We have used simulations to estimate these errors. For these simulations we use a model input angular power spectrum $C^M_{\ell}$ which is  defined as $C^M_{\ell}
=B^{-1} C^{TT}_{\ell}$ (eq. \ref{eq:snr.c}) for   $\ell > \ell^T_m$. In this $\ell$ range we interpolate the estimated $C^{TT}_{\ell}$ to obtain $C^M_{\ell}$ as a continuous function of $\ell$. For  $\ell < \ell^T_m$, we fit a power law to $C^{TT}_{\ell}$ in the vicinity of  $\ell \approx \ell^T_m$ and extrapolate this to calculate $C^M_{\ell}$. 
We construct  several statistically independent realizations of Gaussian random fields corresponding to   $C^M_{\ell}$ and multiply these with the SNR angular profile function $\R$ discussed in the previous section. The images of the simulated SNR were multiplied with the VLA primary beam pattern and used to calculate (eq.~\ref{s1}) the visibilities corresponding to the observed baseline distribution. 

We have assumed that the system noise contribution to both the real and imaginary parts of the visibilities are Gaussian random variables of zero mean and variance $\sigma^2_N$. We have used 
\begin{equation}
\label{sys_noise_formula}
\sigma_N=\frac{SEFD}{\eta_{c}\sqrt{2t_{int}\Delta \nu}}
\end{equation}
where SEFD, $\eta_{c}$ \footnote{https://science.nrao.edu/facilities/vla/docs/manuals/oss/performance/sensitivity} 
are the system equivalent flux density (Jy) and the correlator efficiency respectively, $t_{int}$ is the integration time per visibility in seconds and  $\Delta \nu$ is the channel width in Hz. These values are different for different VLA configurations for a source  and are shown in Table $3$ of \cite{Saha2019}.
The resulting simulated visibilities, which have the simulated SNR signal and simulated system noise,  
were used to determine $C_{\ell}$. We have constructed several statistically independent realizations of these simulations for which the average $C_{\ell}$ matches $C^{TT}_{\ell}$ at $\ell \ge \ell^T_m$.  The variance determined from these  realizations of these simulated  $C_{\ell}$ is used to estimate $\delta C_{\ell}$ for the measured $C^{TT}_{\ell}$. 

 Considering  $C^{AA}_{\ell}$, in addition to the sample variance the  Poisson noise also contributes to the errors. 
 The Poisson noise contribution at any pixel depends on the number of photons that contribute to the value of  $J^A (\bm{\theta})$  at that pixel. Considering the Chandra observation, we have a complicated mapping from the photon counts to  $J^A (\bm{\theta})$ which is carried out by the CIAO software. This poses a difficulty for incorporating the Poisson noise in our analysis, and we have only considered the sample variance contribution to the errors. The sky signal was simulated following exactly the same procedure as for the radio, except that we have used $C^{AA}_{\ell}$ to determine the input model angular power spectrum. The width of the radial profile $\R$ was also set to a value consistent with the X-ray observation. The simulated $J^A (\bm{\theta})$  values were obtained on an image of the same size and resolution as the Chandra image, and these were used to estimate the  simulated $C^{AA}_{\ell}$. We generate $50$ statistically independent realizations of the simulated Chandra images and the variance calculated from these multiple estimates of the simulated $C^{AA}_{\ell}$ were used to compute error $\delta C_{\ell}$ for the estimated $C^{AA}_{\ell}$. The same technique is followed to calculate the error $\delta C_{\ell}$ for the estimated X-ray B $C^{BB}_{\ell}$ and X-ray cross power spectrum $C^{AB}_{\ell}$.

For the radio X-ray cross power spectrum $C^{TA}_{\ell}$, the error variance  for a measurement at a single $\U$ mode is given by
\begin{equation}
    (\delta C^{TA}_{\ell})^2=\frac{1}{2}\,\bigg[(C^{TA}_{\ell})^2+(C^{TT}_{\ell})(C^{AA}_{\ell})\bigg]
    \label{eq:cross-err1}
\end{equation}
where the radio and  X-ray A auto power spectra include the respective signal and the noise contributions. 
The total error variance for an $\ell$ bin  is  also inversely proportional to  the number of independent modes which contribute to the bin.  Here it is difficult to incorporate the individual contributions in the simulations to estimate the errors.  For the present purpose we have made a few simplifying assumptions to estimate these errors.  We assume that the errors $\delta C_{\ell}$ are  primarily dominated by the sample variance which is determined by the number of independent  measurements that contribute to the estimated signal. 
Since the cross-correlation estimator (eq.~\ref{eqn:TGE_V_cross}) has exactly the same sky coverage and baseline distribution as the radio observations, it is then reasonable to assume that  
\begin{equation}
\delta C^{TA}_{\ell}=\frac{C^{TA}_{\ell}}{ C^{TT}_{\ell}} \,\times \,  \delta C^{TT}_{\ell}
\label{eq:cross-err2}
\end{equation}
which we use in this analysis. Similarly, we estimate the errors $\delta C_{\ell}$ for $C^{TB}_{\ell}$ except that $C^{TA}_{\ell}$ is now replaced by $C^{TB}_{\ell}$ in eq. (\ref{eq:cross-err1}) and (\ref{eq:cross-err2}).

\section{Results}
\label{sec:res}
\begin{figure*}
 \begin{center}
  \psfrag{C K}[top]{ $C_\ell$\quad ${\rm K}^2$}  \psfrag{C}[top]{$\ell(\ell+1)\,C_\ell/(2\pi)$ \quad ${\rm K}^2$} \psfrag{l}[bottom]{$\ell$} \psfrag{10}{}
  
   \psfrag{lm3}{$\ell^{T}_m$} \psfrag{lm2}{$\ell^{A}_m$} \psfrag{lm1}{$\ell^{B}_m$}
    \psfrag{lb3}{$\ell^{B}_b$} \psfrag{lb2}{$\ell^{A}_b$}
    \psfrag{lb1}{$\ell^{T}_b$}
  
  \psfrag{ct}{$C^{TT}_{\ell}$}  \psfrag{ca}{$C^{AA}_{\ell}$}  \psfrag{cb}{$C^{BB}_{\ell}$}
  
 \psfrag{2}[bottom]{$10^2$} \psfrag{3}[bottom]{$10^3$} \psfrag{4}[bottom]{$10^4$} \psfrag{5}[bottom]{$10^5$} \psfrag{6}[bottom]{$10^6$} \psfrag{7}[bottom]{$10^7$}
 \psfrag{4}[right]{$10^4$} \psfrag{3}[right]{$10^3$} \psfrag{2}[right]{$10^2$} \psfrag{1}[right]{$10^1$} \psfrag{0}[right]{$10^0$} \psfrag{-1}[right]{$10^{-1}$} \psfrag{-2}[right]{$10^{-2}$} \psfrag{-3}[right]{$10^{-3}$} \psfrag{-4}[right]{$10^{-4}$} \psfrag{-5}[right]{$10^{-5}$} \psfrag{-6}[right]{$10^{-6}$} \psfrag{-7}[right]{$10^{-7}$} \psfrag{-8}[right]{$10^{-8}$} \psfrag{-9}[right]{$10^{-9}$} \psfrag{-10}[right]{$10^{-10}$} \psfrag{-11}[right]{$10^{-11}$} \psfrag{-12}[right]{$10^{-12}$} \psfrag{-13}[right]{} \psfrag{-14}[right]{$10^{-14}$} \psfrag{-15}[right]{} \psfrag{-16}[right]{$10^{-16}$} \psfrag{-17}[right]{} \psfrag{-18}[right]{$10^{-18}$}\psfrag{-19}[right]{} \psfrag{-20}[right]{$10^{-20}$}
  \includegraphics[scale=0.29,angle=-90]{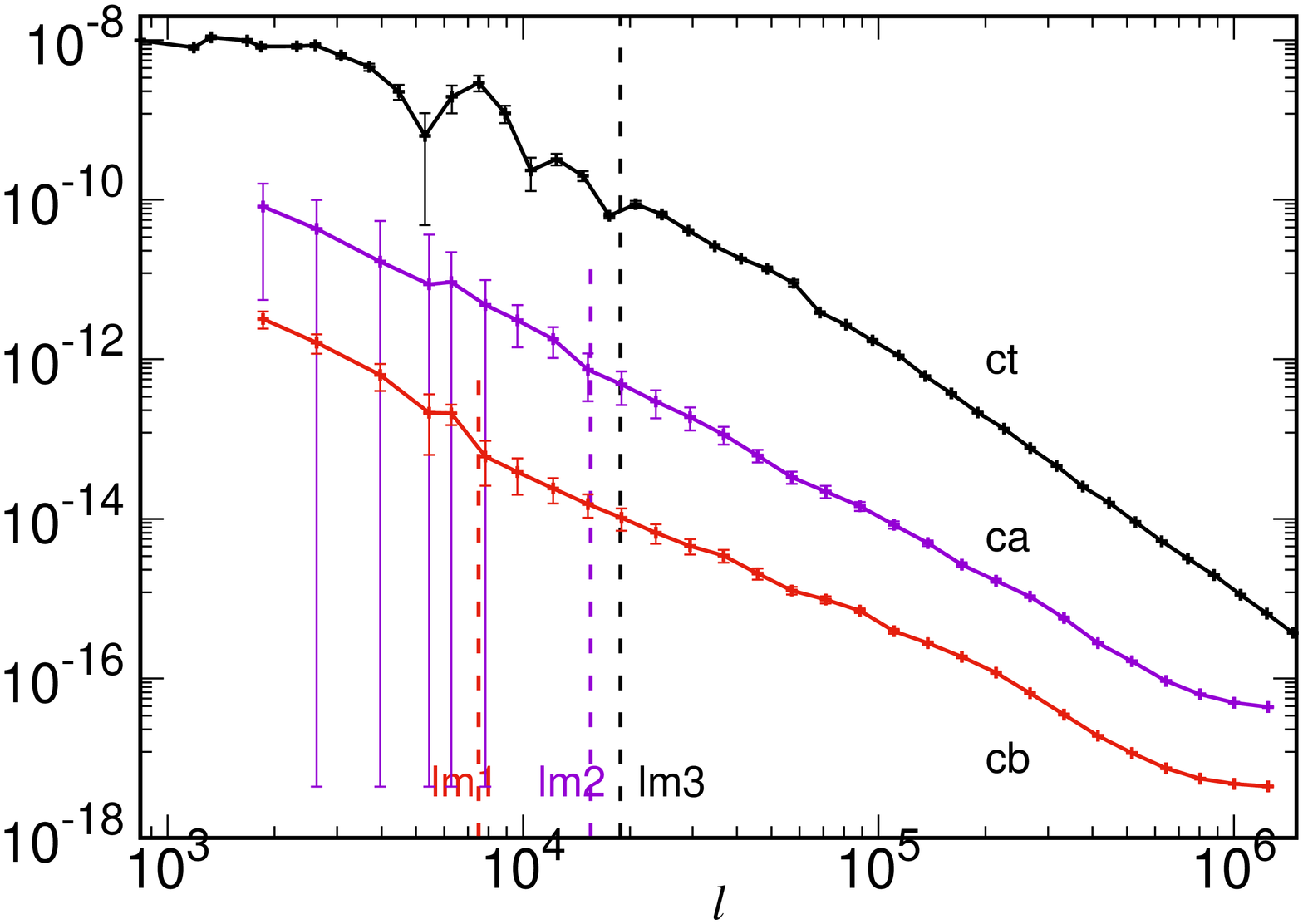}  \hspace{0.8cm}\includegraphics[scale=0.29,angle=-90]{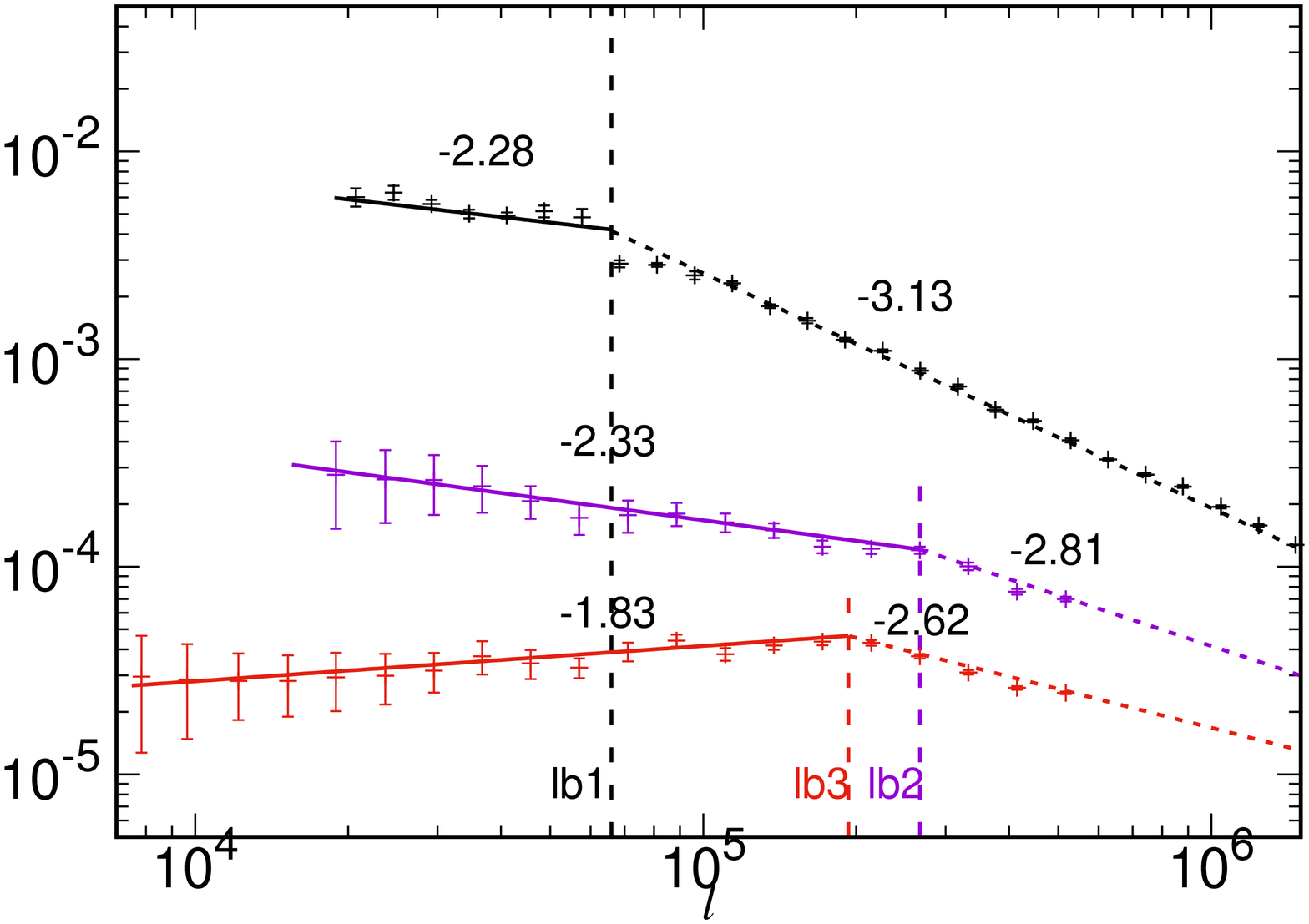}
 \caption{ Left panel: Estimated angular auto power spectra $C^{TT}_{\ell}$, $C^{AA}_{\ell}$ and $C^{BB}_{\ell}$
  with their respective $1-\sigma$ error estimates. Right panel: The scaled angular auto power spectra for the $\ell$ range used for fitting,  along with the respective best fit broken power-law. The values of the power law index $\beta$ are shown. These values and the break $\ell_b$ are also provided in Table $\ref{tab:CasApsfit}$.}
 \label{autops}
 \end{center} 
\end{figure*}

We first consider the three power spectra $C^{TT}_{\ell}$, $C^{AA}_{\ell}$ and $C^{BB}_{\ell}$ which are shown in Figure $\ref{autops}$.   
Throughout the rest of the paper, in addition to $C_{\ell}$
 we have shown  the scaled angular power spectrum $\ell(\ell+1)\,C_{\ell}/2\pi$. For radio we 
 can interpret $\ell(\ell+1)\,C^{TT}_{\ell}/2\pi$  as the variance of the observed brightness temperature fluctuations at the  angular scale corresponding to $\ell$, the other power spectra may be similarly interpreted in terms of photon number counts, etc. Considering  $C^{TT}_{\ell}$ first,  the $\ell$ range shown in the left panel  corresponds to the entire baseline range of the radio observations.
The $1-\sigma$  error bars shown in the figure are quite large at small $\ell$ where the sample variance   dominates.  
The error bars are  smaller at large $\ell$ where they  are not noticeable in the figure. 
We see that  $C^{TT}_{\ell}$ flattens out at $\ell \leq \ell_m^T (=1.88\times10^4)$. This flattening can be attributed to the convolution (eq.~\ref{eq:snr.b}) due to the finite angular extent  of the SNR.
We may interpret  $C^{TT}_{\ell}$  in  the $\ell$ range  $\ell > \ell_m^T$ as arising from specific intensity fluctuations within the SNR, and we restrict the subsequent analysis to this $\ell$ range. 
We see that in this $\ell$ range  $C^{TT}_{\ell}$ shows a power law like behaviour, however a single power law  does not provide a good fit for the entire $\ell$ range as there is a  break at  $\ell^T_b =6.60 \times 10^{4}$. We find that separately fitting two different power laws of the form $C_{\ell}=A \, \ell^{\beta}$, one for $\ell < \ell^T_b$ and another for $\ell > \ell^T_b$  provides a reasonably good fit which is shown in the right panel of Figure $\ref{autops}$.   We obtain the best fit 
power law index $\beta=-2.28\pm 0.08$  and $\beta=-3.13\pm 0.01$ for $\ell < \ell^T_b$ and $\ell > \ell^T_b$ respectively. The power law index steepens by $-0.85$ as we go across the break from small $\ell$ to large $\ell$. 
The goodness of fit and the fitting parameters are summarized in Table \ref{tab:CasApsfit}.
We have also tried fitting a   broken power law  with $\ell^T_b$ as an additional  parameter, however this  does not improve the reduced $\chi^2$. These findings have been  reported previously in \cite{Saha2019}. We  note that an earlier work 
\citep{roy09} has analyzed the same visibility data to estimate $C^{TT}_{\ell}$ for the  Cas A SNR, however the power spectrum estimator used there was  different. 
The earlier  work also found a break at  $\ell^T_b \sim 6.60\times10^4$, with a power law having   $\beta=-2.22\pm0.03$ at the small  $\ell$ values in   the  range  $1.00\times10^4-6.28\times10^4$ and a different power law having 
 $\beta=-3.23\pm0.09$ in the $\ell$ range $\ell >6.28\times10^4$. We find that our results here are  consistent with the results reported in the earlier work.

\begin{table*}
 \centering
 \begin{tabular}{|c|c|c|c|c|c|c|c|}
          		\hline
band    & $\ell_m$ & $\ell_b $ & power index $\beta$ & power index $\beta$ & $\ell$ range  &  No.of points & $\frac{\chi^2}{DOF}$\\
& & &  $(\ell \le \ell_b)$ & $(\ell > \ell_b)$ & for fitting & & \\ 
       \hline
radio T & $1.88\times 10^{4}$ &  $ 6.60\times 10^{4}$  & $-2.28\pm 0.08$  & & $ \ell_m - \ell_b$  & $7$ & $0.86$\\ 
 & &  &   & $-3.13\pm 0.01$ & $8.17\times 10^{4}-1.88 \times 10^{6}$ & $18$ & $3.44$ \\ \hline
 X-ray A & $1.55\times 10^{4}$  & $(2.67\pm0.36)\times 10^{5}$   	 &  $-2.33\pm 0.05$ & $-2.81\pm 0.10$ & $\ell_m -  6.00 \times10^5$   & 16 & 0.89   \\ 
 \hline
 X-ray B & $7.50\times 10^{3}$ & $(1.93\pm0.11)\times 10^{5}$   &  $-1.83\pm 0.05$ & $-2.62\pm 0.03$ &  $\ell_m - 6.00\times10^5$ 	&  $20$ & $1.23$  \\ 
 \hline
 cross TA & $1.80\times10^4$ & $(2.67\pm0.43)\times 10^{5}$ &  $-2.27\pm 0.04$ &  $-2.83\pm 0.17$ & $\ell_m - 6.00\times10^5$ & $20$ & $9.17$ \\ \hline
    cross TB & $1.80\times 10^{4}$ &    $(1.16\pm0.08)\times10^5$  &  $-2.11\pm 0.05$ &  $-2.76\pm 0.03$  & $\ell_m - 6.00 \times 10^{5}$  &   $20$ & $3.69$ \\  \hline
    cross AB &	$7.50\times 10^{3}$ & $(2.11\pm0.23)\times 10^{5}$ &  $-2.11\pm0.06$ & $-2.70\pm0.04$ & $\ell_m - 6.00 \times 10^{5}$ &$20$ & $2.32$ \\ 
              \hline
 \end{tabular}
 \caption{The values of the parameters obtained by fitting $C^{TT}_{\ell}$, $C^{AA}_{\ell}$, $C^{BB}_{\ell}$, $C^{TA}_{\ell}$, $C^{TB}_{\ell}$ and $C^{AB}_{\ell}$  of Cas A SNR. }
 \label{tab:CasApsfit}
 \end{table*}

\begin{figure}
 
 \psfrag{s b}[top]{$\R$ \quad [arbitary units]} \psfrag{radius}[bottom]{$\tv$ (arcmin)} 
 \psfrag{ 0}[bottom]{$0$}  \psfrag{ 1}[bottom]{$1$}  \psfrag{ 2}[bottom]{$2$}  \psfrag{ 3}[bottom]{$3$}  \psfrag{ 0.5}[bottom]{$0.5$}
 \psfrag{ 1.5}[bottom]{$1.5$} \psfrag{ 2.5}[bottom]{$2.5$} \psfrag{ 3.5}[bottom]{$3.5$}
 
  \psfrag{-200}{$-200$}  \psfrag{ 0}[right]{0}  \psfrag{ 0.02}{$0.02$}  \psfrag{ 0.04}{$0.04$}  \psfrag{ 0.06}{$0.06$}  \psfrag{ 0.08}{$0.08$}  \psfrag{ 0.1}{$0.1$}  \psfrag{ 0.12}{$0.12$}  \psfrag{ 0.14}{$0.14$}  \psfrag{ 1600}[right]{}  \psfrag{ 1800}{$1800$} 
 \includegraphics[width=0.32\textwidth,scale=0.29,angle=-90]{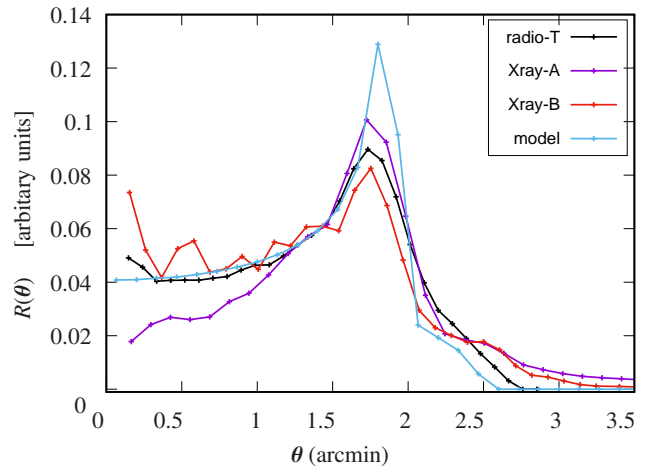}	
 \vspace{0.5cm}
 \caption{The normalised radio and X-ray radial profiles $\R$ estimated from the respective images in Figure \ref{fig:AIPSCasA}. The 2D profile for the simulation model used in Section \ref{sec:disc} is also shown.}
 \label{fig:radial_prof}

\end{figure}

 We next consider the X-ray A angular power spectrum  $C^{AA}_{\ell}$ shown in the left panel of  Figure $\ref{autops}$. We see that  $C^{AA}_{\ell}$ flattens out for $\ell\leq \ell^A_m(=1.55\times10^4)$ whereas it is a declining function of $\ell$ for $\ell > \ell^A_m$.  Similar to radio, we attribute  this flattening  to the convolution (eq.~\ref{eq:snr.b}) due to the finite angular extent  of the SNR. We see that the value of $\ell^A_m$ is close to the value of $\ell^T_m=1.88\times 10^4$ obtained for radio which indicates that  the  Cas A SNR is expected to have the same angular extent in radio and X-ray A (Figure \ref{fig:AIPSCasA}). To verify this we consider the respective radial profiles $\R$  shown in Figure $\ref{fig:radial_prof}$. These have been determined  by averaging the intensity along various radii emanating from the centers of the  respective  images in Figure $\ref{fig:AIPSCasA}$. The radial profiles are all normalised using $\int \R \, d^2 \theta =1$.  We find that the radial profiles  of the radio and X-ray A images both peak at $\theta \approx 1.8 ^{'}$ confirming that they have   very similar  angular extents. 
 Note that the angular extent estimated from Figure $\ref{fig:radial_prof}$ is possibly larger than the values predicted by one-dimensional self-similar hydrodynamics models \citep{HamiltonSarazin1984,LamingHwang2003,HwangLaming2012} due to projection effects. 
 We may interpret  $C^{TT}_{\ell}$  in  the $\ell$ range  $\ell > \ell_m^T$ as arising from specific intensity fluctuations within the SNR, and we restrict the subsequent analysis to this $\ell$ range.  As can be seen from Figure $\ref{autops}$, the error estimates are large at $\ell <9.00\times 10^4$ due to the large sample variance,  the error decreases thereafter. From the left panel of Figure $\ref{autops}$, $C^{AA}_{\ell}$ appears to hit a floor at the three largest $\ell$ values. We have omitted these data points from the subsequent  analysis
 which is restricted to the $\ell$ range $\ell^A_m - 6.00\times 10^5$.  Like radio, here also a single power law does not provide a good fit to the data.  We have obtained a good fit using a broken power law considering the position of the break $\ell^A_b$ as a free parameter. The best fit values and the goodness of fit are summarized in Table~\ref{tab:CasApsfit}. We obtain $\ell^A_b=  (2.67\pm0.36)\times10^5$ with  the power law index values  $\beta=-2.33\pm 0.05$ and $\beta=-2.81\pm 0.10$ for $\ell<\ell^A_b$ and $\ell>\ell^A_b$ respectively.  The power law slope $\beta$ steepens by $-0.48$ across $\ell^A_b$ from low to high $\ell$.

We next consider the X-ray B angular power spectrum  $C^{BB}_{\ell}$ shown in the left panel of  Figure $\ref{autops}$. Similar to $C^{TT}_{\ell}$ and $C^{AA}_{\ell}$, we see that $C^{BB}_{\ell}$ also flattens out at $\ell \leq \ell_m^B(=7.50\times10^3)$.  The values of $\ell_m^T$ for radio and $\ell_m^A$ for X-ray A are respectively  $2.5$ and $2.1$ times larger than $\ell_m^B$ for X-ray B. Taking  this at face value,  this indicates that the angular extent of the SNR in the radio C Band and X-ray A are approximately $2.5$ and $2.1$ times smaller than that in  X-ray B respectively. By comparing the three radial profiles (Figure \ref{fig:radial_prof}), and also by directly comparing the  images in  Figures \ref{fig:AIPSCasA},   we see that the angular extents are nearly the same in radio, X-ray A and X-ray B. 
The origin of the difference between the values of  $\ell_m^T$, $\ell_m^A$ and $\ell_m^B$  is not clear at present and is discussed in Section \ref{sec:disc}.  Considering  $C^{BB}_{\ell}$ in Figure  $\ref{autops}$, we see  that the estimated error bars of $C^{BB}_{\ell}$ are quite large at smaller $\ell$ ($\ell<5.00\times 10^4$) where we have a relatively large sample variance, the errors  decreases at larger  $\ell$. Like $C^{AA}_{\ell}$, $C^{BB}_{\ell}$  appears to hit a floor  at  the  four largest $\ell$ values (left panel of Figure $\ref{autops}$). 
To understand the origin of this floor, we have simulated images of Cas A using the radial profile function $\R$ (Figure \ref{fig:radial_prof}) and $C^{BB}_{\ell}$ as the input and having mean similar to that of the observed X-ray B image. Next, we have applied a detailed Chandra instrument model (marx, \cite{marx}) to these simulated images to incorporate the instrumental effects like point spread function and Poisson noise from a finite number of photons. We find that both the effects affect the angular power spectra computed from these images only at large $\ell$ i.e. $\ell > 6.00\times 10^{5}$ where it appear flatter than the angular power spectra calculated from the simulated images before the application of marx. For $\ell>6.00\times 10^{5}$ corresponding to angular scales less than $2.16^{\arcsec}$, the point spread function $(\sim 0.5^{\arcsec}-3^{\arcsec})$ away from the center can smooth out spatial structures. We have excluded this range from the subsequent analysis.
We have used $\chi^2$ minimisation to determine the best fit  power law  for $C^{BB}_{\ell}$  in the $\ell$ range 
  $\ell^B_m-6.00\times 10^{5}$. Like $C^{TT}_{\ell}$ and $C^{AA}_{\ell}$, we find that a single power law does not provide a good fit to the data.  We have obtained a good fit to  the data 
  using a broken power law  with the position of the break $\ell^B_b$ also included  as a free parameter.
  We have used this to identify  a break at  $\ell^B_b=(1.93\pm0.11)\times 10^{5}$.  
  The fit to the data is shown in the right panel of Figure $\ref{autops}$, and the values of the slopes and the reduced $\chi^2$ are shown in Table \ref{tab:CasApsfit}.  We have obtained the slopes 
  $\beta=-1.83\pm 0.05$  and $\beta=-2.62\pm 0.03$  for $\ell < \ell^B_b$ and $\ell > \ell^B_b$ respectively. 
  We find that the power law index steepens by $-0.79$ as we go across the break from small $\ell$ to large $\ell$.

\begin{figure*}

    \psfrag{C K}[top]{$\ell(\ell+1)C_{\ell}/(2\pi)$ \quad ${\rm K}^2$} \psfrag{ l}{} \psfrag{l}[bottom]{$\ell$}  \psfrag{lb2}{$\ell^{AB}_b$}  \psfrag{lb1}{$\ell^{TA}_b$} \psfrag{lmab}{$\ell^{AB}_m$} \psfrag{lb}{$\ell^{TB}_b$} \psfrag{lmta}{$\ell^{TA}_m$} \psfrag{lmtb}{$\ell^{TB}_m$}
  \psfrag{10}{} \psfrag{TA}{$C^{TA}_\ell$} \psfrag{TB}{$C^{TB}_\ell$} \psfrag{AB}{$C^{AB}_\ell$}
 \psfrag{2}[bottom]{$10^2$} \psfrag{3}[bottom]{$10^3$} \psfrag{4}[bottom]{$10^4$} \psfrag{5}[bottom]{$10^5$} \psfrag{6}[bottom]{$10^6$} \psfrag{7}[bottom]{$10^7$}
 \psfrag{6}[right]{$10^6$} \psfrag{5}[right]{$10^5$}\psfrag{4}[right]{$10^4$} \psfrag{3}[right]{$10^3$} \psfrag{2}[right]{$10^2$} \psfrag{1}[right]{$10^1$} \psfrag{0}[right]{$10^0$} \psfrag{-1}[right]{$10^{-1}$} \psfrag{-2}[right]{$10^{-2}$} \psfrag{-3}[right]{$10^{-3}$} \psfrag{-4}[right]{$10^{-4}$} \psfrag{-5}[right]{$10^{-5}$} \psfrag{-6}[right]{$10^{-6}$} \psfrag{-7}{} \psfrag{-8}[right]{$10^{-8}$} \psfrag{-9}{} \psfrag{-10}[right]{$10^{-10}$} \psfrag{-11}{} \psfrag{-12}[right]{$10^{-12}$} \psfrag{-13}[right]{} \psfrag{-14}[right]{$10^{-14}$} \psfrag{-15}[right]{} \psfrag{-16}[right]{$10^{-16}$}
 \includegraphics[scale=0.2,angle=-90]{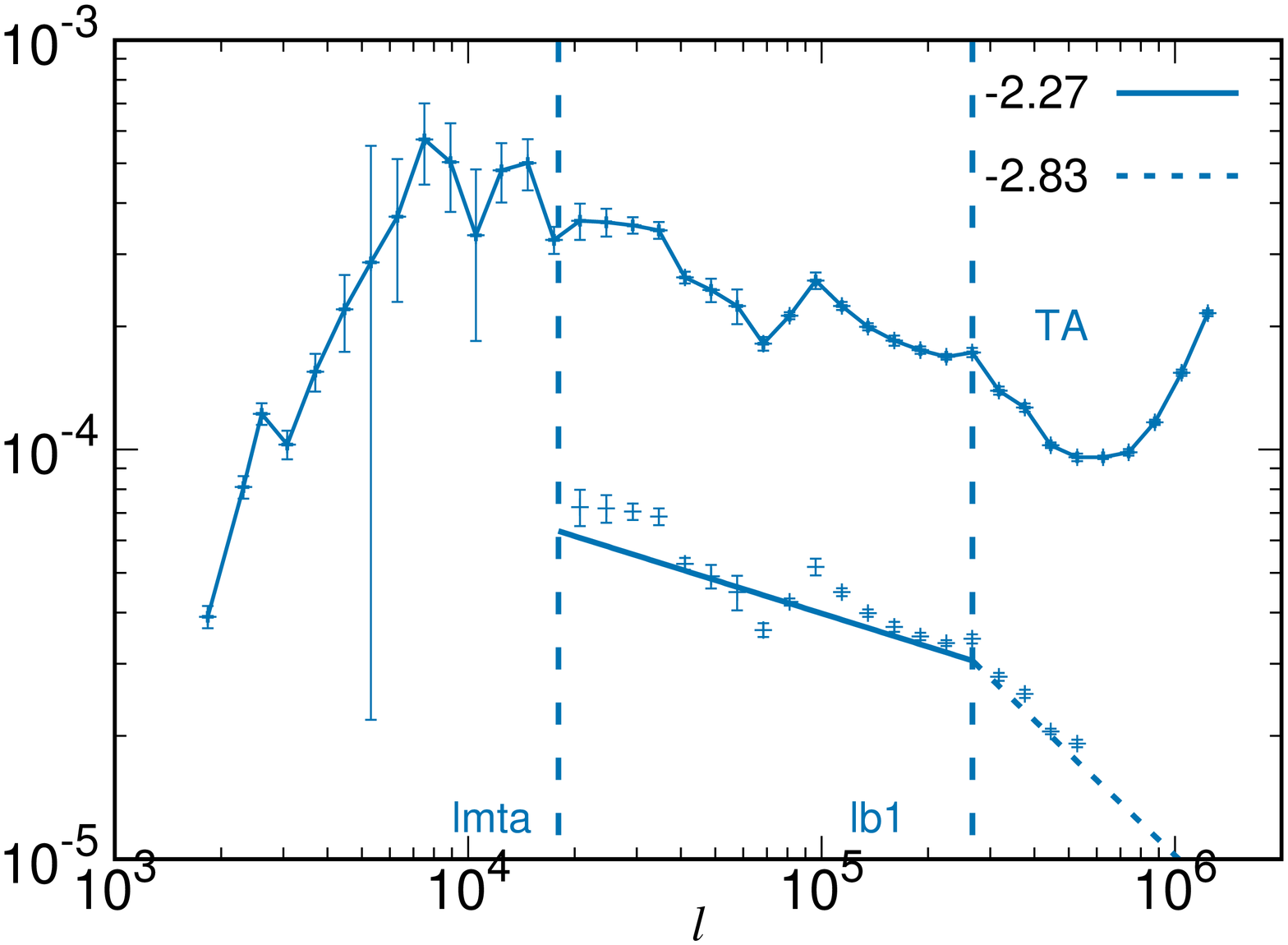} \hspace{0.4cm}\includegraphics[scale=0.2,angle=-90]{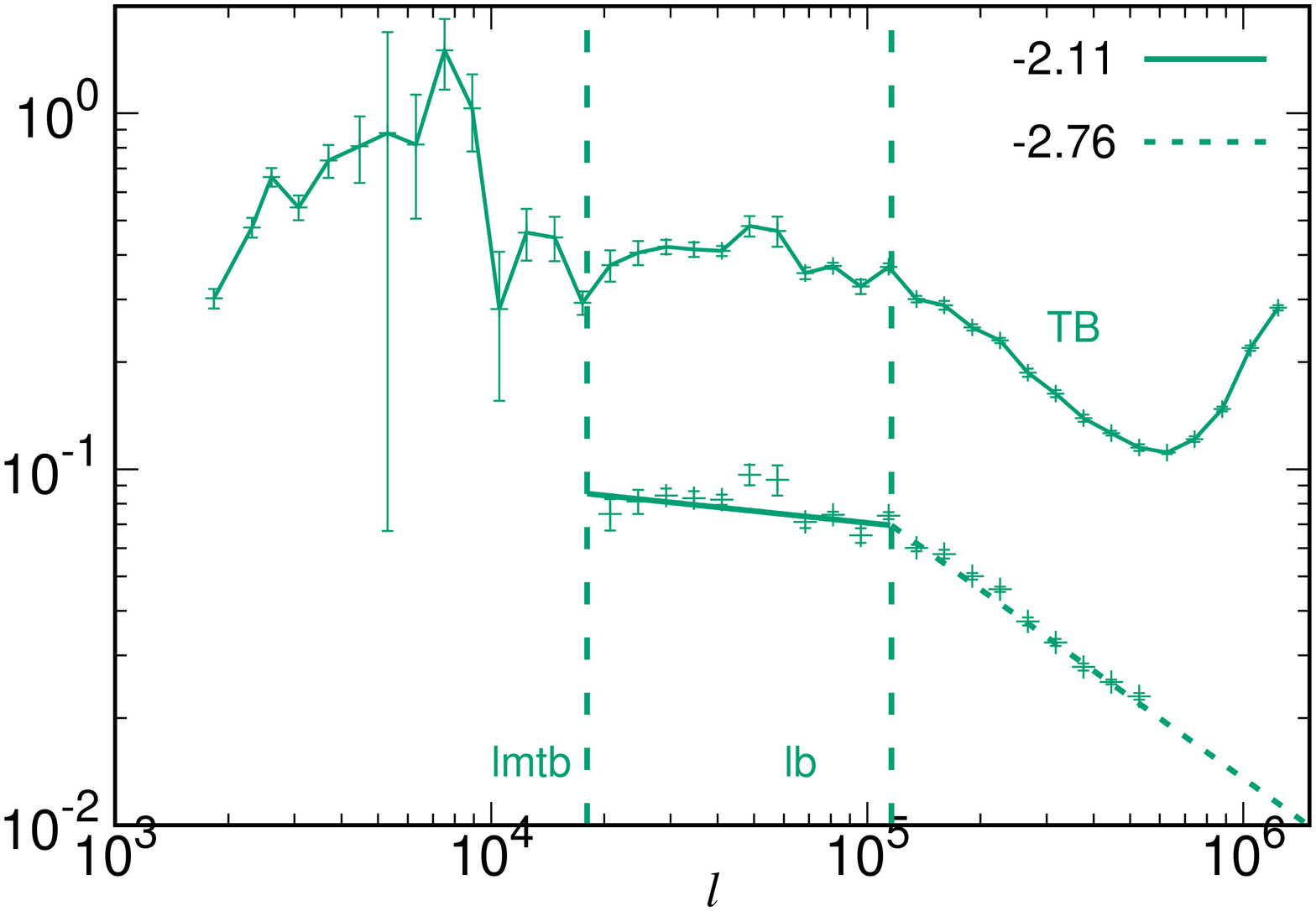}\hspace{0.4cm} \includegraphics[scale=0.2,angle=-90]{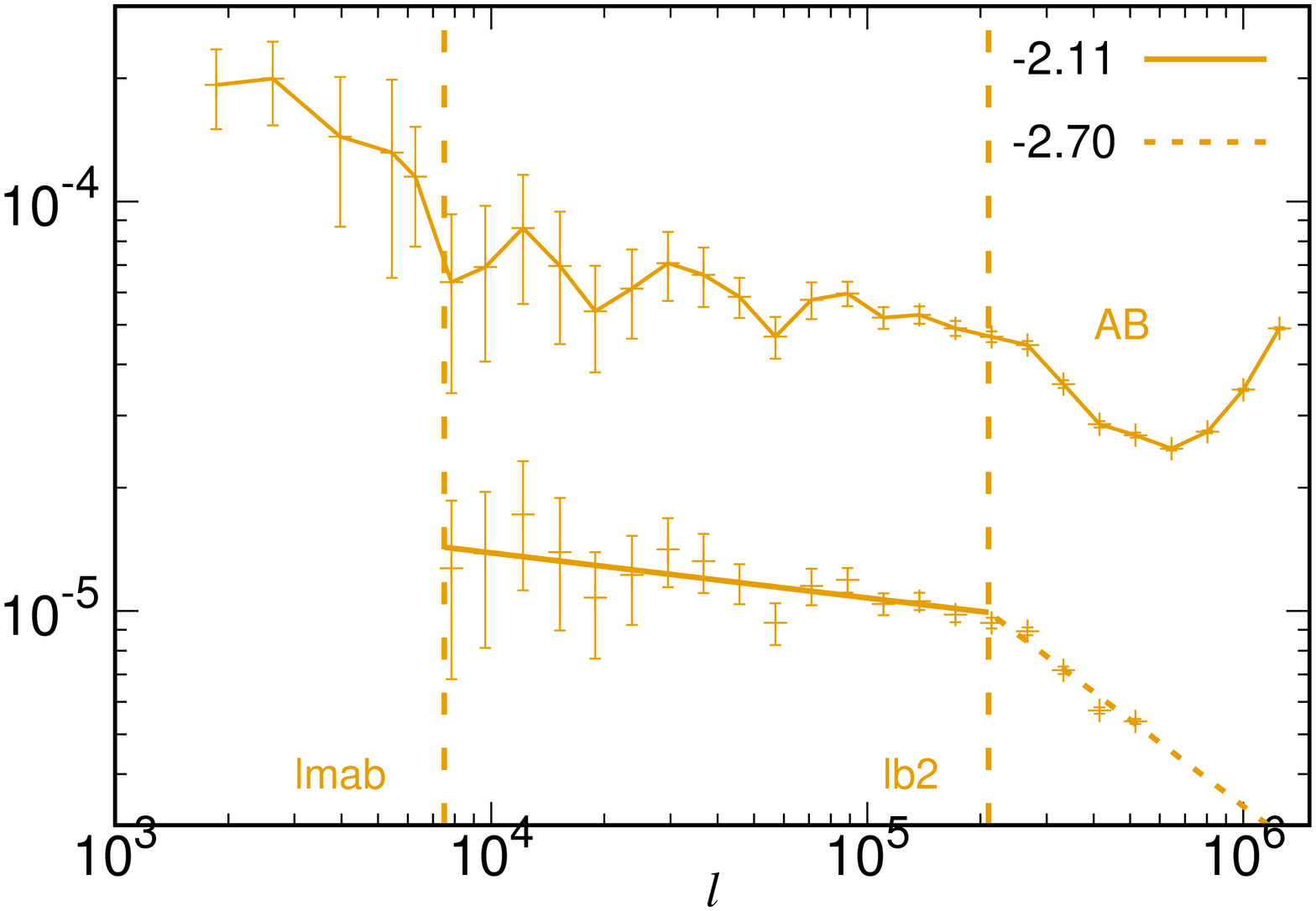}	
 \caption{Cross  power spectra  $C_{\ell}^{TA}$ (left); $C_{\ell}^{TB}$ (center); $C_{\ell}^{AB}$ (right)
scaled with $\ell(\ell+1)/(2\pi)$  along with $1-\sigma$ error bars. Respective best fit  broken power laws are also shown with an offset in amplitude.} 
\label{fig:cross2}
\end{figure*}

 For the radio X-ray cross power spectra  $C^{TA}_{\ell}$ and $C^{TB}_{\ell}$ the respective  X-ray  data were  converted to have exactly the same baseline distribution as the radio data, and the binned $\ell$ values exactly match those of $C^{TT}_{\ell}$. The left and middle panels of Figure \ref{fig:cross2}
show $C^{TA}_{\ell}$ and $C^{TB}_{\ell}$ respectively.  We find that $C^{TA}_{\ell}$ and $C^{TB}_{\ell}$ 
both flatten out due to the convolution (eq.~\ref{eq:snr.b}) at  small $\ell$ below 
$\ell^{TA}_m=\ell^{TB}_m=1.80\times 10^4$. This value  matches that of $\ell^{T}_m$ and is also comparable to $\ell^A_m$.   Similar to $C^{AA}_{\ell}$ and $C^{BB}_{\ell}$,  for both $C^{TA}_{\ell}$  and  $C^{TB}_{\ell}$ we observe the presence of a floor at  large $\ell$. The occurrence of this flat region (floor) in both  $C^{TA}_{\ell}$ and $C^{TB}_{\ell}$ indicates that this feature is not due to the X-ray photon Poisson noise.The Poisson noise due to the finite number of X-ray photons is not expected to be correlated with the radio signal. For both $C^{TA}_{\ell}$ and $C^{TB}_{\ell}$ we have  restricted the subsequent fitting to the $\ell$ range  $ \ell^{TA}_m-6.00\times10^5$.
A single power law does not provide a good fit for any of these two power spectra. Considering a broken power law with the position of the break as an additional free parameter, for 
 $C^{TA}_{\ell}$ we obtain the best fit parameters $\beta=-2.27\pm 0.04$ ($\ell \le \ell^{TA}_b$) and  $\beta=-2.83\pm 0.17$ ($\ell > \ell^{TA}_b$) with $\ell^{TA}_b=(2.67\pm0.43)\times 10^{5}$.
 We however note that this  does not provide a good fit to $C_{\ell}^{TA}$ as there appears to be several breaks and different slopes in the $\ell$ range   $ \ell^{TA}_m - \ell^{TA}_b$.
 Considering $C^{TB}_{\ell}$, we have obtained the break at  $\ell^{TB}_b=(1.16\pm0.08)\times10^5$ with slopes  $\beta=-2.11\pm 0.05$ and $\beta=-2.76\pm 0.03$ for $\ell < \ell^{TB}_b$ and  $\ell > \ell^{TB}_b$ respectively. The best fit curves for $C^{TA}_{\ell}$ and $C^{TB}_{\ell}$ 
 are shown in the respective panels of Figure \ref{fig:cross2}, whereas the goodness of fit and the best fit parameters are summarized in Table~\ref{tab:CasApsfit}.  We note that our errors for $C_{\ell}^{TB}$ are possibly underestimated leading to a  $\chi^2$ which is somewhat on the higher side.

 The right panel of Figure \ref{fig:cross2} shows  the cross power spectrum $C^{AB}_\ell$ between X-ray A and B evaluated using   eq. (\ref{eq:xc-ps}).  We see that $C^{AB}_\ell$ becomes flat for  $\ell \leq \ell^{AB}_m(=7.50 \times 10^3)$, this  value is similar to $\ell^B_m$. The estimated error bars are quite large for  $\ell<7.00\times10^4$, thereafter the errors decrease at larger $\ell$.  Like the other X-ray auto and cross power spectra,  $C^{AB}_\ell$ too hits  a floor at the largest  $\ell$ values. We have restricted the $\ell$ range to $\ell^{AB}_m-6.00\times 10^5$ for subsequent fitting. We find that a single power law does not provide  a good fit across the entire  $\ell$ range considered here. We have fitted $C^{AB}_{\ell}$ with a  broken power law with the position of the break $\ell^{AB}_b$ as a free parameter. The best fit broken power law is also shown in the right panel of Figure $\ref{fig:cross2}$, with an offset in the amplitude for clarity. We have obtained the break at  $\ell^{AB}_b=(2.11\pm0.23)\times 10^{5}$ with slopes  $\beta=-2.11\pm 0.06$ and $\beta=-2.70\pm0.04$ for $\ell < \ell^{AB}_b$ and  $\ell > \ell^{AB}_b$ respectively. The best fit parameters are presented in   Table~\ref{tab:CasApsfit}.

\begin{figure*}
\minipage{0.33\textwidth}
 \psfrag{10}{} \psfrag{corr coeff}{$c^{TA}_{\ell}(\times10^{-4})$}  \psfrag{2}[bottom]{$10^2$} \psfrag{3}[bottom]{$10^3$} \psfrag{4}[bottom]{$10^4$} \psfrag{5}[bottom]{$10^5$} \psfrag{6}[bottom]{$10^6$} \psfrag{7}[bottom]{$10^7$}
     \psfrag{ 0}[right]{0}  \psfrag{0.00010}{$1.0$}  \psfrag{ 0.00045}[right]{}\psfrag{0.00020}{$2.0$}  \psfrag{0.00018}[right]{}  \psfrag{0.00016}[right]{}  \psfrag{0.00014}[right]{}   \psfrag{0.00012}[right]{} \psfrag{0.00030}{$3.0$}   \psfrag{0.00022}[right]{}  \psfrag{0.00024}[right]{} \psfrag{0.00026}[right]{}  \psfrag{0.00028}[right]{}  \psfrag{ 0.7}{$0.7$}  \psfrag{0.00008}[right]{}  \psfrag{ 0.8}{$0.8$}  \psfrag{ 0.85}[right]{}  \psfrag{ 0.9}{$0.9$} \psfrag{ 1.0}{$1.0$} \psfrag{ 0.95}{} \psfrag{l}[bottom]{$\ell$}
 
 \includegraphics[scale=0.2,angle=-90]{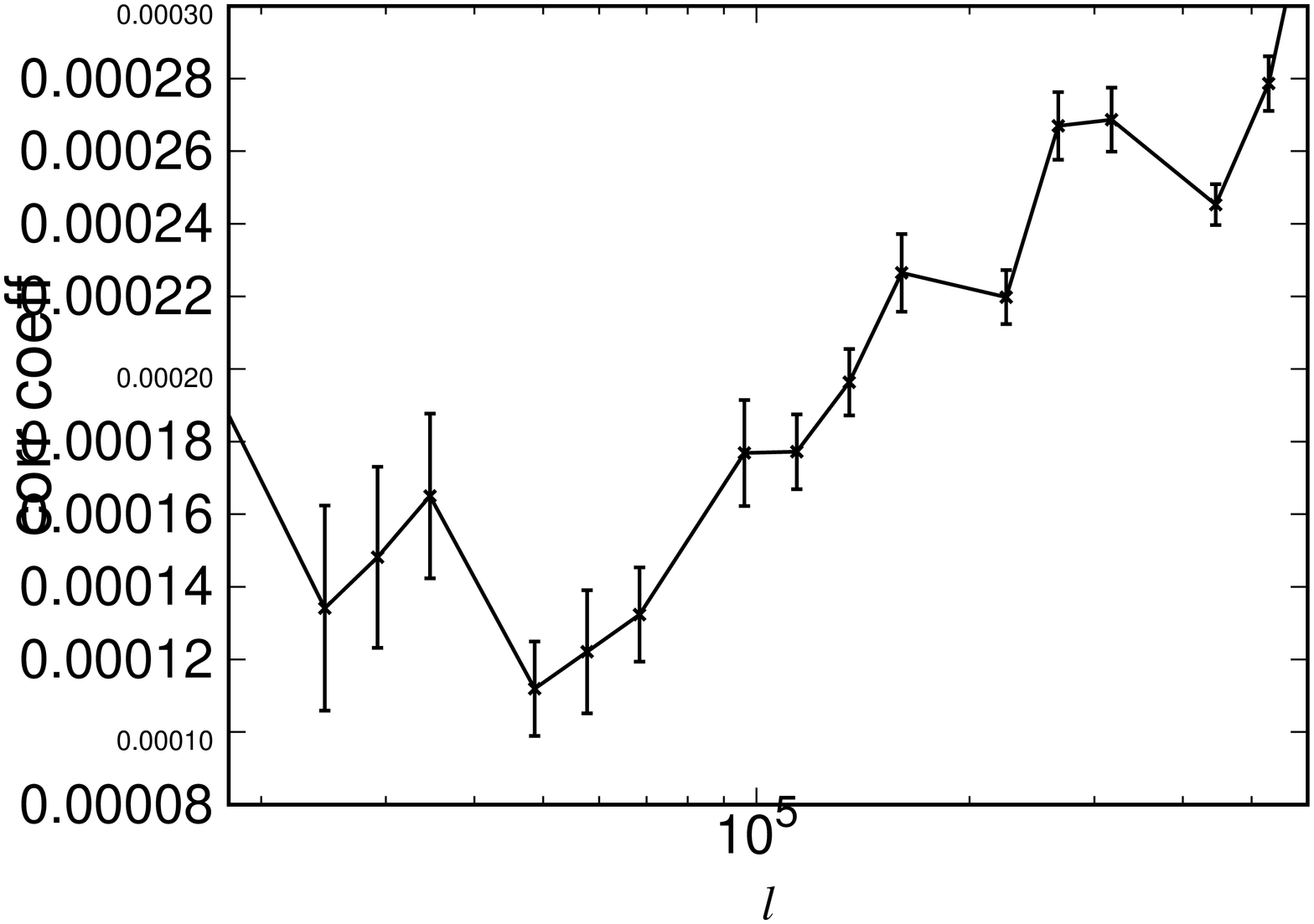} 	
\endminipage\hfill
\minipage{0.33\textwidth}
 \psfrag{10}{} \psfrag{corr coeff}{$c^{TB}_{\ell}$}
     \psfrag{2}[bottom]{$10^2$} \psfrag{3}[bottom]{$10^3$} \psfrag{4}[bottom]{$10^4$} \psfrag{5}[bottom]{$10^5$} \psfrag{6}[bottom]{$10^6$} \psfrag{7}[bottom]{$10^7$}
     \psfrag{ 0}[right]{0}  \psfrag{ 0.4}{$0.4$}  \psfrag{ 0.45}[right]{}\psfrag{ 0.5}{$0.5$}  \psfrag{ 0.55}[right]{}  \psfrag{ 0.6}{$0.6$}  \psfrag{ 0.65}[right]{}  \psfrag{ 0.7}{$0.7$}  \psfrag{ 0.35}[right]{}  \psfrag{ 0.8}{$0.8$}  \psfrag{ 0.85}[right]{}  \psfrag{ 0.9}{$0.9$} \psfrag{ 1.0}{$1.0$} \psfrag{ 0.95}{} \psfrag{l}[bottom]{$\ell$}
 \includegraphics[scale=0.2,angle=-90]{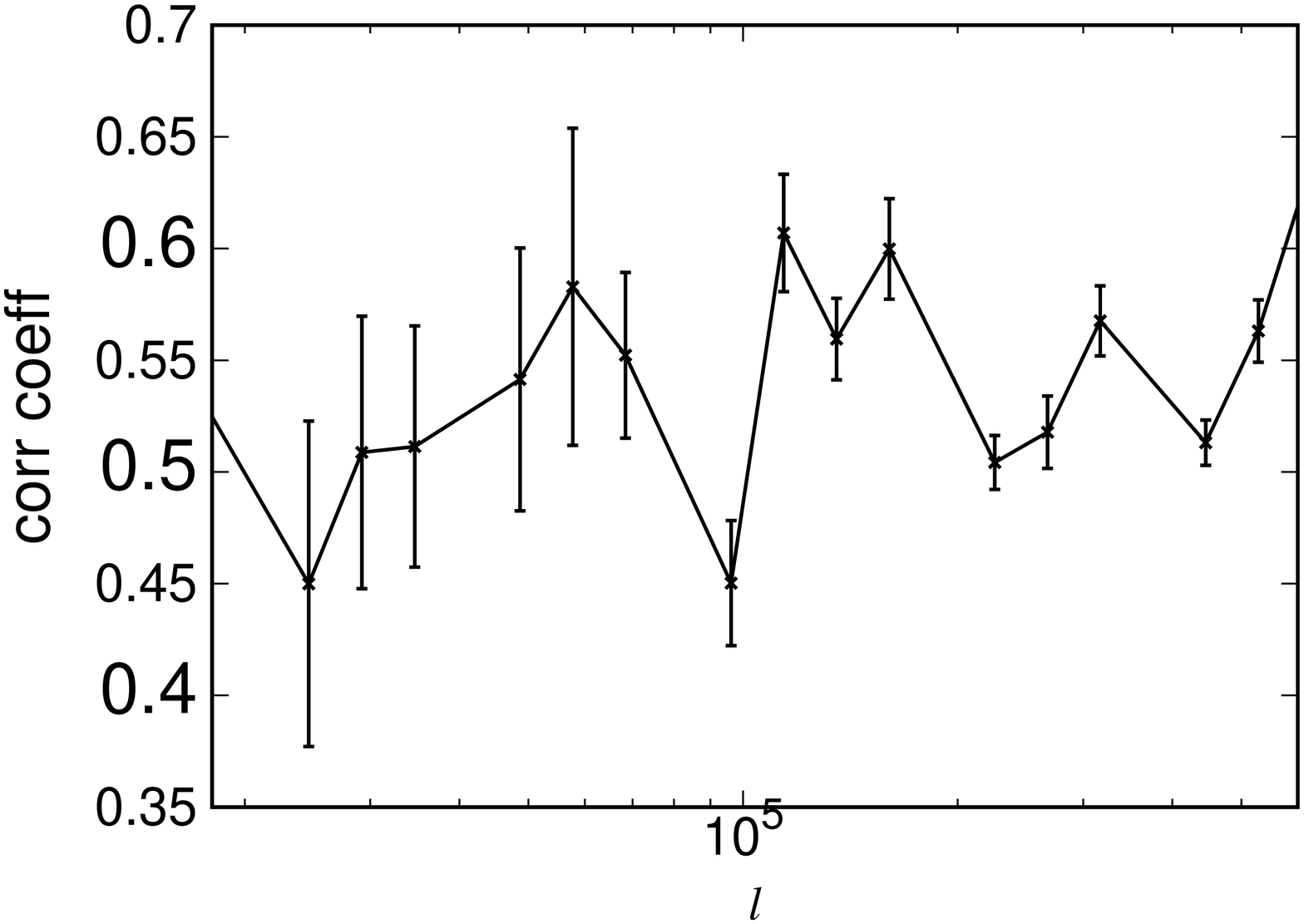} 	
\endminipage\hfill
\minipage{0.33\textwidth}
 \psfrag{10}{} \psfrag{corr coeff}{$c^{AB}_{\ell}$}
    \psfrag{2}[bottom]{$10^2$} \psfrag{3}[bottom]{$10^3$} \psfrag{4}[bottom]{$10^4$} \psfrag{5}[bottom]{$10^5$} \psfrag{6}[bottom]{$10^6$} \psfrag{7}[bottom]{$10^7$}
     \psfrag{ 0}[right]{0}  \psfrag{ 0.2}{$0.2$}  \psfrag{ 0.3}[right]{}  \psfrag{ 0.4}{$0.4$}  \psfrag{ 0.5}[right]{}  \psfrag{ 0.6}{$0.6$}  \psfrag{ 0.7}[right]{}  \psfrag{ 0.8}{$0.8$}  \psfrag{ 0.9}[right]{}   \psfrag{ 1}[right]{$1.0$} \psfrag{ 0.1}{} \psfrag{l}[bottom]{$\ell$}
 \includegraphics[scale=0.2,angle=-90]{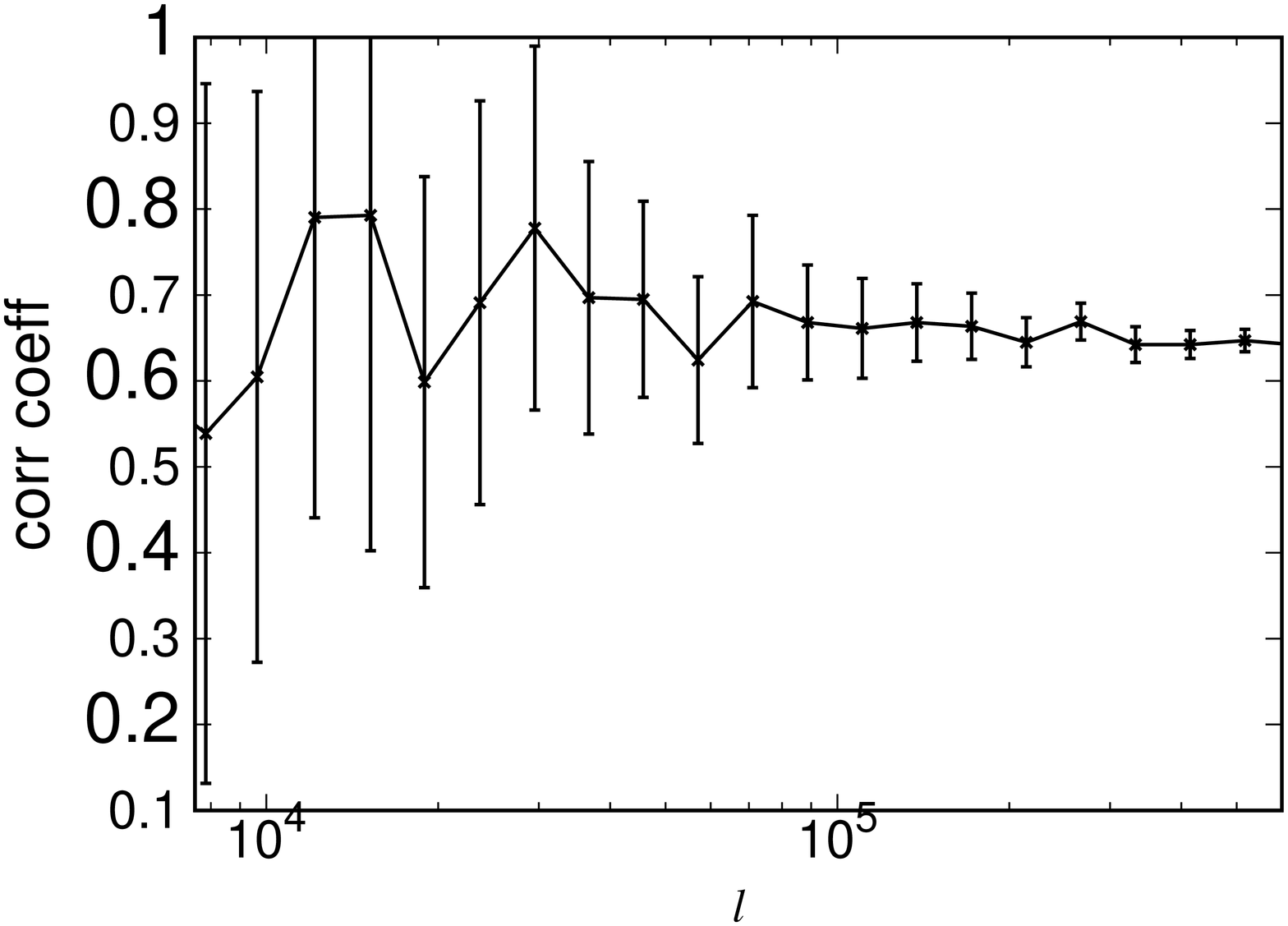}

\endminipage\hfill
 \vspace{0.4cm}
 \caption{Correlation coefficient ($c_{\ell}$) for the cross power spectra  $C_{\ell}^{TA}$ (left);  $C_{\ell}^{TB}$ (center);  $C_{\ell}^{AB}$  (right).  }
 \label{fig:corr_coeff}
\end{figure*}

We next consider the dimensionless correlation coefficient  
\begin{equation}
    c^{AB}_{\ell}=\frac{C^{AB}_{\ell}}{\sqrt{C^{AA}_{\ell}\,C^{BB}_{\ell}}}
\label{eq:cc}
\end{equation} 
which, for a fixed $\ell$, quantifies the strength of the correlation between  the angular fluctuations in X-ray A with those  in X-ray B. 
The correlation coefficients $c^{TA}_{\ell}$ and  $c^{TB}_{\ell}$ are similarly defined. 
The correlation coefficients are   expected to have  values in the range $-1$ to $+1$, with the value $1$ occurring if the two signals are perfectly correlated. We expect the correlation to have a values $0$ and $-1$ if the two signals are uncorrelated 
and anti-correlated respectively. 
The left, middle and right panels of Figure $\ref{fig:corr_coeff}$ respectively show the  correlation coefficients $c^{TA}_{\ell}$, $c^{TB}_{\ell}$ and $c^{AB}_{\ell}$. The  $1-\sigma$ error bars shown in the respective panels of Figure $\ref{fig:corr_coeff}$   account for the errors in both the cross-power spectra as well as the auto-power spectra. 
Considering $c^{TA}_{\ell}$, we find that the values are positive but very small ($ \sim 10^{-4}$). In contrast, both $c^{TB}_{\ell}$ and $c^{AB}_{\ell}$ also have positive values but the values are relatively large  $(0.4 - 0.8)$, with the values of $c^{AB}_{\ell}$ being slightly larger than those of $c^{TB}_{\ell}$. This shows that X-ray B is highly correlated with both radio and X-ray A. However, X-ray A has extremely weak correlations with radio. 
The  radio frequency  emission from SNRs like Cas A  is predominantly non-thermal synchrotron radiation. For Cas A,  X-ray A  is known to be predominantly   thermal emission \citep{Vink96,willingale02}. The fact that radio and X-ray A are very weakly correlated bears out the picture  that these two originate from two different emission mechanisms. Interestingly, X-ray B is correlated with  both  radio and X-ray A, this supports  a picture where X-ray B is a mixture of both thermal and non-thermal emission.

Table $\ref{tab:CasApsfit}$ summarizes the best fit parameters obtained for the different  estimated power spectra. We find that  all the estimated power spectra, except $C^{TA}_{\ell}$,    can be represented by a broken power law with a break at $\ell_b$ where the power law slope $\beta$ has a particular value for $\ell\le\ell_b$ and a different value for  $\ell >\ell_b$.  In all cases the slope is negative ($\beta <0$), and it steepens  (change $\Delta \beta <0$) from small $\ell$ to large $\ell$ across the break. 
Considering the auto-power spectra, for radio, X-ray A and B respectively  the  position of the break has  values $\ell_b=(0.66,2.67,1.93) \times 10^5$  with $\beta =-2.28,-2.33,-1.83$ for $\ell \le \ell_b$ and 
$\beta =-3.13,-2.81,-2.62$  for $ \ell > \ell_b$. We see that for $\ell \le \ell_b$ X-ray A has the steepest power spectrum whereas radio is steepest for $ \ell > \ell_b$ and also overall. Considering the cross-power spectra, we exclude $C^{TA}_{\ell}$ where the correlation is extremely weak and the broken power law does not provide a good fit. For $C^{TB}_{\ell}$ and $C^{AB}_{\ell}$ the positions of the respective breaks are $\ell_b=(1.16, 2.11) \times 10^5 $ with $\beta=-2.11,-2.11$ for $\ell \le \ell_b$ and $\beta=-2.76,-2.70$ for $\ell > \ell_b$. Considering the change in slope, we have $\Delta \beta=-0.85,-0.48, -0.79,-0.65$ and $-0.59$ for $C^{TT}_{\ell}, C^{AA}_{\ell}, C^{BB}_{\ell}, C^{TB}_{\ell}$  and $C^{AB}_{\ell}$ respectively. 
Here we see that radio is somewhat unique as the position of the break $\ell^{TT}_b$ is smallest,   the slope $\beta=-3.13$ is the steepest and the change in slope $\Delta \beta=-0.85$ is the largest among all the power spectra estimated here. 
For X-ray B the slopes are approximately $0.5$ less steeper than radio and the change in slope $\Delta \beta=-0.79$ is comparable to that for radio. For X-ray A the slope is slightly steeper than  radio for $\ell \le \ell_b$, however the change $\Delta \beta$ is much smaller.  Considering the cross-correlation $C^{TB}_{\ell}$, we see that $\ell^{TB}_b$ is the geometric mean of $\ell^{T}_b$ and $\ell^{B}_b$, while the value of $\beta$ is close to that of radio for $\ell \le \ell^{TB}_{\ell}$ and  it is close to that of X-ray B for  $\ell > \ell^{TB}_{\ell}$.  Considering the cross-correlation $C^{AB}_{\ell}$, we see that  
$\ell^{AB}_b$ is slightly smaller than the mean of $\ell^{A}_b$ and $\ell^{B}_b$, however the values of $\beta$ are very close to those for $C^{TB}_{\ell}$.

\section{Modelling the radio power spectrum $C^{TT}_{\ell}$}
\label{sec:model_radio}
Considering the  angular power spectrum $C^{TT}_\ell$ of the Cas A SNR in radio, we interpret this 
 as arising from fluctuations  in the synchrotron radiation which in turn trace the  fluctuations in the electron density and the  magnetic field  both  of which results from  MHD turbulence in the ionized plasma of the SNR. 
 It is likely that the contribution from the fluctuations in the electron density is subdominant to the contribution from the fluctuations in the magnetic field. 
 For $\ell >\ell^{T}_m$ we find that  $C^{TT}_\ell$  is a broken power law with 
  a steepening of  the power law index from $\beta=-2.28$ for $\ell^{T}_m \leq \ell<\ell^{T}_b$  to $\beta=-3.13$ for  $\ell \geq \ell^{T}_b$.  
  Several earlier  works  \citep{roy09,Saha2019} have interpreted  such a  change in the power law index in terms of a transition from 2D to 3D turbulence at an angular scale   of $\theta^{T}_b \approx \pi/\ell_b^{T}$, which turns out to have  a value $\theta^{T}_b = 0.16\arcmin$ for  Cas A.  The angular scale $\theta^{T}_b$ can be associated with the shell thickness of Cas A. 
   Recently, \cite{Choudhuri2020} have also shown that the angular scale of the break of the observed power spectrum for a shell-type SNR is related to its shell thickness. Using simulations, they present a systematic study of the effect of shell geometry and the projection of SNR on to a 2D plane which can cause the turbulence to change from 3D to 2D at angular scales greater than the shell thickness. 
  At scales smaller than the shell thickness, the SNR shell can have Fourier modes of fluctuations in three independent spatial directions. However at scales greater than the shell thickness, there will be no modes perpendicular to the shell thickness.  At these length-scales the SNR shell can have Fourier modes of fluctuations in only two independent spatial directions. We expect this transition to be manifested as a break in the power spectrum.  Considering incompressible  
  3D Kolmogorov turbulence \citep{kolmogorov1941}, the energy spectrum follows a power law with slope ${-5/3}$. The equivalent velocity power spectrum is predicted to have slopes  $-8/3$ and $-11/3$  for  2D and 3D turbulence respectively \citep{Frisch95, Elmegreen2004}.  Further,  
  the density power spectrum is also predicted to  follow the velocity power spectrum
  \citep{GS95}. The values $\beta=-2.28$ and  $\beta=-3.13$ and a difference of slope 
  $\Delta \beta =-0.85$  obtained here for Cas~A  are roughly consistent with these predictions. These findings  strongly indicate that the measured radio power spectrum originates from MHD turbulence in the plasma  of the SNR, and the break is associated with the geometrical effect of the shell thickness of Cas A. It is interesting to note that a similar break is also found in the  the angular power spectrum of HI emission from the disk of external spiral galaxies. The  location of the break there is associated with the  scale height of the disk, and this has been used to estimate the scale height of the galaxy NGC $628$ \citep{Dutta2008}.

We have carried out 3D simulations where we have implemented a  simple geometrical model to illustrate  the  above interpretation of the observed $C^{TT}_\ell$ of Cas A.  The simulations are in terms of 3D angular unit $\mathbf{r}$ and its Fourier conjugate $\mathbf{k}$ for which $|\mathbf{k}|=\ell$.  We have generated  Gaussian random   brightness temperature fluctuations $\delta T(\mathbf{r})$ inside a 3D cube of size $[20.48^{'}]^3$ with $[1024]^3$ grid points and spacing $0.02^{'}$. 
  The simulated  $\delta T(\mathbf{r})$ corresponds to a 3D power spectrum $P(k)=Ak^{-3.1}$ which matches the observed  $C^{TT}_\ell$ at small angular scales or large $\ell$.
   To replicate the emission structure as seen from the radio observation (left panel of Figure \ref{fig:AIPSCasA}), we first introduce a single spherical shell of outer radius $2.5\arcmin$ and thickness $0.16\arcmin$ which is the value of $\theta^{T}_b$. The shell radius is approximately the observed size of the remnant. Here, there are two regions, the shell and the core which is a sphere enclosed within the inner radius of the shell. The simulated fluctuations $\delta T(\mathbf{r})$ only fill up the shell,  we have erased the fluctuations outside the shell as well as  inside the core. The resulting cube is projected to a 2D plane which corresponds to the plane of the sky. We have estimated the angular power spectrum from the projected 2D brightness temperature fluctuations. We find that the simulated $C_\ell$ matches the observed $C^{TT}_{\ell}$ at large $\ell$, and the position of the break also  is  close to the observed $\ell_b^{T}$. 
   However, we find that at $\ell < \ell_b^{T}$ the  simulated $C_{\ell}$ is considerably steeper  that the observed $C^{TT}_{\ell}$ and the simulated amplitude also is larger.  The position of the break can be shifted by adjusting the shell thickness, however we find that the mismatch in slope and amplitude persists even when we vary the shell thickness in the range $0.125\arcmin-0.25\arcmin$. 
   
\begin{figure}
 \psfrag{f}[right]{$f(\mathbf{r})$} \psfrag{r}[bottom]{$\mathbf{r}$ (arcmin)}
 \psfrag{ 0}[bottom]{$0$}  \psfrag{ 1}[bottom]{$1.0$}  \psfrag{ 2}[bottom]{$2.0$}  \psfrag{ 3}[bottom]{$3.0$}  \psfrag{ 0.5}[bottom]{$0.5$}
 \psfrag{ 1.5}[bottom]{$1.5$} \psfrag{ 2.5}[bottom]{$2.5$} \psfrag{ 3.5}[bottom]{$3.5$}
  \psfrag{ 0}{$0$}  \psfrag{ 0.2}{$0.2$}  \psfrag{ 0.4}{$0.4$} \psfrag{ 0.6}{$0.6$} \psfrag{ 0.8}{$0.8$}
  \psfrag{ 1.0}{$1.0$} \psfrag{ 1.2}{$1.2$}
 \includegraphics[width=0.32\textwidth,scale=0.29, angle=-90]{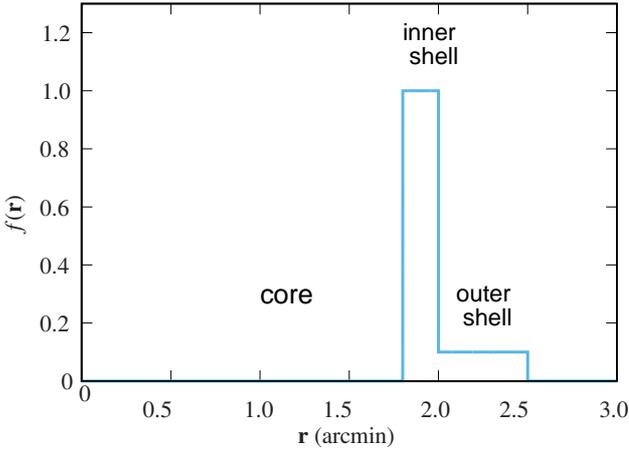}	
 \hspace{15cm}
 \vspace{0.2cm}
 \caption{The 3D radial profile used to simulate the double shelled SNR structure considered here.} 
\label{fig:3dstruc}

\end{figure}

     It is interesting to note that the radio image in Figure \ref{fig:AIPSCasA} actually hints at a double shelled structure with a  bright inner shell corresponding to the shocked ejecta and a fainter outer shell corresponding to the shocked CSM.  In order to emulate this  we have tried out different  double spherical shell structures, here we focus on one for which the simulated $C_{\ell}$ is closest to the   observed $C^{TT}_{\ell}$. 
   Figure $\ref{fig:3dstruc}$ shows the radial structure of the 3D double spherical shell 
   consisting of  an inner shell of radius $2.0\arcmin$ and thickness $0.2\arcmin$ and an outer spherical shell of radius $2.5\arcmin$ and thickness $0.5\arcmin$. We now have three distinct simulated regions namely  the empty core, the inner shell and the outer shell. 
    The simulated brightness temperature fluctuations fill the inner shell while the same fluctuations with reduced amplitude fill up the outer shell. The fluctuations inside the core and outside the outer shell are erased. Figure \ref{fig:radial_prof} also shows the radial profile $\R$ corresponding to the 2D projected 
    image of the double shelled structure. We see that the simulated $\R$ closely matches the observed radio profile  interior to the peak which occurs at $\approx 1.8 ^{'}$, however the simulated peak amplitude is larger and the simulated profile drops faster at larger angles as compared to the observed profile.  
    
\begin{figure}
 
   \psfrag{scaled C K}[top]{$\ell(\ell+1)\,C_\ell/2\pi$ \quad ${\rm K}^2$} \psfrag{ l}{} \psfrag{l}[bottom]{$\ell$} \psfrag{Cband}{$C^{TT}_{\ell}$} \psfrag{model}{$C_{\ell}$}
  \psfrag{10}{} 
 \psfrag{2}[bottom]{$10^2$} \psfrag{3}[bottom]{$10^3$} \psfrag{4}[bottom]{$10^4$} \psfrag{5}[bottom]{$10^5$} \psfrag{6}[bottom]{$10^6$} \psfrag{7}[bottom]{$10^7$}
 \psfrag{5}[right]{$10^5$} \psfrag{4}[right]{$10^4$} \psfrag{3}[right]{$10^3$} \psfrag{2}[right]{$10^2$} \psfrag{1}[right]{$10^1$} \psfrag{0}[right]{$10^0$} \psfrag{-1}[right]{$10^{-1}$} \psfrag{-2}[right]{$10^{-2}$} \psfrag{-3}{} \psfrag{-4}[right]{$10^{-4}$} \psfrag{-5}{} \psfrag{-6}[right]{$10^{-6}$} \psfrag{-7}{} \psfrag{-8}[right]{$10^{-8}$} \psfrag{-9}{} \psfrag{-10}[right]{$10^{-10}$} \psfrag{-11}{} \psfrag{-12}[right]{$10^{-12}$}
  \includegraphics[width=0.32\textwidth,scale=0.29,angle=-90]{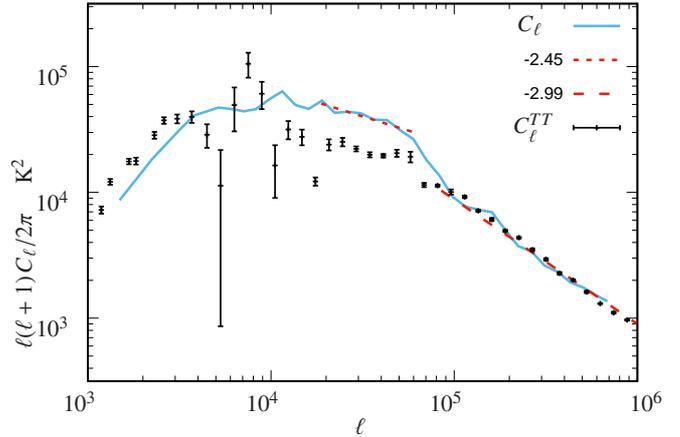}
 \caption{Comparison of  the observed $C^{TT}_{\ell}$ with the 
 model  $C_{\ell}$ obtained from the simulations. The red dotted and dashed lines respectively show the best fit power laws to the model  $C_{\ell}$ in two different $\ell$ ranges.} 
 \label{CasApscomp}
\end{figure}

     We have used several statistically independent realizations of the simulated $\delta T(\mathbf{r})$  to obtain the mean $C_{\ell}$ shown in Figure \ref{CasApscomp}.  We find that  simulated power spectrum $C_{\ell}$ can be fitted with two different power laws in two different $\ell$ ranges, one  with  
     $\beta=-2.45\pm0.06$ in the range $\ell=(2.0-7.0)\times10^4$  and another with $\beta=-2.99\pm0.04$ in the range $\ell=8.5\times10^4-7.0\times10^5$. At large $\ell$, the slope $\beta=-2.99$ is roughly consistent (by construction)  with the observed slope of $\beta=-3.13$. At small $ \ell$ the slope  $\beta=-2.45$ is still somewhat steeper than the measured value of $\beta=-2.28$, and the simulated amplitude also is larger. However, we note that the double shelled simulation is closer to the measured $C^{TT}_{\ell}$ than is possible with a single shelled structure. The inner shell thickness $0.2\arcmin$ is reflected in the break at $\ell \sim7.00\times 10^4$  which is close to the measured value $\ell^{T}_b=6.60\times10^4$. Although the model considered here is not able to accurately reproduce the observed $C^{TT}_{\ell}$, the similarity between the simulated $C_{\ell}$ and  the observed $C^{TT}_{\ell}$ may be interpreted as indicating that what we have observed is indeed a geometrical effect associated with the shell thickness. The difference between the two may be attributed to the possibility  that the actual SNR  structure may be more complex than the double shelled structure considered here. The fact that the peak in the observed $\R$ has a smaller amplitude and larger angular extent at $\theta > 2^{'}$ as compared to the simulation  ( Figure \ref{fig:radial_prof}) also points to the same possibility. A better match may be possible with more detailed modelling of the 3D shell structure, however this is beyond the scope of the present paper and we do not attempt this here.

\vspace{0.5cm}

  The double cascade theory  of 2D turbulence proposed by \cite{Kraichnan1967, kraichnan_1971} provides an  alternative explanation for the broken power law power spectrum observed here.  This  theory predicts an inverse energy cascade with  an energy spectrum of slope  ${-5/3}$ on  length-scales larger than the length-scale where energy is injected into the system,  and a direct enstrophy cascade with  an energy spectrum of  slope  ${-3}$ at smaller  length-scales. Several subsequent studies (e.g. \citealt{lindborg99}, \cite{Boffetta2010}) have validated this scenario for 2D turbulence. Here we find that  the corresponding slopes of $-8/3$ and $-4$ predicted for the 2D velocity power spectrum  roughly match the  slopes $\beta=-2.28$  and $\beta=-3.13$  obtained  for Cas A. Knot-shock interactions in the CSM provides a possible mechanism for  energy and vorticity injection into  the  SNR plasma \citep{Beresnyak_2009}  at a length-scale which is comparable to the Cas A shell thickness. The validation of the above interpretation requires detailed simulations which we intend to address in  future work.

\section{Discussion \& Conclusion}
\label{sec:disc}
 
  Synchrotron radiation is the dominant  emission from SNRs at radio frequencies. We have interpreted the 
 radio angular power spectrum $C^{TT}_\ell$ of Cas A SNR  as arising from fluctuations of MHD turbulence in the synchrotron emitting plasma.  It is known that the synchrotron emissivity depends on the electron number density and the magnetic field component perpendicular to the line of sight. The estimated  $C^{TT}_\ell$ is expected to quantify the underlying fluctuations of both the electron density and the magnetic field. $C^{TT}_\ell$ shows a broken power law with the power law index $\beta=-2.28$ at $\ell < \ell^T_b =6.60\times10^4$ steepening to $\beta=-3.13$ at $\ell \geq \ell^T_b$. 
 The values of power law indices and the amount of steepening of the power law slope $\Delta 
\beta=-0.85$ from small $\ell$ to large $\ell$ are consistent with the predictions of incompressible Kolmogorov turbulence \citep{kolmogorov1941}.
Like in earlier works \citep{roy09, Saha2019}, we have identified the 
angular scale corresponding to the break  $\theta^{T}_b \approx \pi/\ell_b^{T} = 0.16\arcmin$ with the shell thickness of Cas A  which sets the transition scale from 2D to 3D turbulence. We have carried out simulations to validate the above interpretation. We find that a double shelled structure is  able to qualitatively match some of the features in the observed $C^{TT}_\ell$, however quantitative differences persist between the simulated and observed $C^{TT}_\ell$. The rough match indicates that the break in the power spectrum is possibly a geometrical effect, however a more complex shell structure  is needed for the model to quantitatively reproduce the observed $C^{TT}_\ell$.
Moreover, in a companion paper \citep{Choudhuri2020},
detailed simulations are carried out to study various effects such as remnant's shell thickness, projection from $3$D structure onto a $2$D observational plane, presence of diffuse foreground emission, on the observed power spectrum of a SNR.

We now discuss the X-ray properties of Cas A. The spectral characteristics of Cas A in X-ray varies on all angular scales 
 due to fluctuations in the  underlying radiation processes and composition \citep{Hughes2000}, 
 down to the resolution limit of the telescope which corresponds to physical scales of $0.02$ pc. Considering the global spectrum shown in   Figure \ref{fig:xspec}, we see that this matches  \cite{heldervink08}. Further, both X-ray A and B considered here are devoid of any strong line emission. At X-ray frequencies, the emission from SNRs can be broadly of two types, the thermal emission which usually refers to a combination of bremsstrahlung and line emission and the non-thermal emission which usually refers to X-ray synchrotron emission. 
 For both radio and X-ray, the synchrotron emissivity depends on the electron density and the magnetic field component perpendicular to the line of sight. 
 The fluctuations in the synchrotron radiation are probably caused mainly by fluctuations in the background magnetic field. 
 In contrast, the  bremsstrahlung  emissivity depends on  the product of the electron and ion densities \citep{rybicki&lightman}. 
 Moreover, these density fluctuations are related to the magnetic field fluctuations by flux freezing. 
  The energy $0.5-10$ keV band which include both X-ray A and B, can be well fitted with a thermal model \citep{Vink96,willingale02}. However, for X-ray B, a part of the continuum emission is non-thermal synchrotron \citep{Vink99} and the contribution of synchrotron was found to be less than one-third \citep{Vink&Laming03}. 
  Later using 1Ms data of Cas A SNR of \cite{Hwang2004}, \cite{heldervink08} claimed that non-thermal emission accounts for $54\%$ of the total emission of X-ray B while the rest is thermal bremsstrahlung. Interestingly, the size and the width of the forward shock and the inner ring of the X-ray continuum reported in \cite{heldervink08} are roughly similar to the value of the radius and shell thickness of outer and inner shell respectively used in the shell structure simulations that we have conducted in this paper.
 Particularly for X-ray B, we understand that it is difficult to completely disentangle thermal from non-thermal component, even with the best data. This is in contrast with X-ray A which is inferred to be predominantly thermal.
  Furthermore accounting for the non-thermal bremsstrahlung, \cite{Vink08} showed that for Cas A it is restricted to very high energies above  $100$ keV which is much greater than X-ray A and B.

 Here we now discuss the estimated X-ray power spectra $C^{AA}_\ell$ and  $C^{BB}_\ell$.
Like the radio power spectrum  $C^{TT}_\ell$, we also model both $C^{AA}_\ell$ and $C^{BB}_\ell$ as  broken power laws in the $\ell$ range $\ell_m-6.00\times10^5$. The close match between  $\ell^A_m=1.55\times10^4$ and  $\ell^T_m=1.88 \times 10^4$ indicates similar angular extents for the Cas A SNR in  radio and X-ray A, whereas $\ell^B_m=7.5 \times 10^3$ indicates a larger   angular extent (nearly twice) in X-ray B. However, comparing the various  radial profiles  (Figure \ref{fig:radial_prof}) we see that $\R$ peaks at roughly the same value $\tv \approx 1.8\arcmin$ indicating very similar angular extents in radio, X-ray A and B. However, the peak amplitude differs with  X-ray A, radio and X-ray B occurring in decreasing order.  The width of the peak is similar for all three, whereas at $\tv < 1.2\arcmin$ 
the amplitude of $\R$ for X-ray A  is smaller than radio and X-ray B which have comparable amplitude. 
We however note that for X-ray B $\R$ shows considerable fluctuations which indicates that the sources responsible for this emission have a much more clumpy distribution as compared to the radio and X-ray A. Considering $C_{\ell}^{AA}$, for $\ell<\ell^A_b=2.67\times10^5$ we have $\beta=-2.33$  which is slightly steeper (by $-0.05$) as compared to $C^{TT}_\ell$,   whereas for $\ell\ge\ell^A_b$ we have 
 $\beta=-2.81$ which is shallower than $C^{TT}_\ell$  by $0.32$. Considering $C^{BB}_\ell$, we have    $\beta=-1.83$ for $\ell<\ell^{B}_b=1.93\times 10^5$ and $\beta=-2.62$  for   $\ell > \ell^{B}_b$,    which are both shallower by approximately $0.5$ as compared to the corresponding slopes for $C^{TT}_\ell$. Comparing the two X-ray power spectra we see that the slope of $C^{BB}_{\ell}$ is shallower than $C^{AA}_{\ell}$ by approximately $0.5$ and $0.2$ at $\ell < \ell_b$ and $\ell > \ell_b$ respectively. 
 We note that $\ell^{T}_b < \ell^{B}_b <\ell^{A}_b$. In the earlier discussion, for radio we have associated the break in the power law as arising from a transition from 3D to 2D turbulence at an angular scale comparable to the SNR shell thickness.  If we interpret the various $\ell_b$  in terms of the respective shell thicknesses, we then  have the largest shell thickness  $\theta^{T}_b = 0.16\arcmin$ followed by $\theta^{B}_b = 0.056\arcmin$ and $\theta^{A}_b = 0.040 \arcmin$ respectively. The analysis in Section \ref{sec:model_radio} indicates that such an interpretation may be valid for radio and also X-ray B both of which have $\Delta \beta \approx -0.8$ which is comparable to $\Delta \beta=-1$ expected in a transition from 2D to 3D. However for X-ray A we have $\Delta \beta=-0.48$ which is considerably smaller,  and this is unlikely to arise from purely geometrical considerations.

The radio, X-ray A and B signals all originate from the same turbulent plasma of the SNR, and we can expect the angular fluctuations in these three different signals to be correlated. However, each may   originate from a different emission mechanism and therefore     trace a different aspect of the fluctuations in the turbulent plasma and possibly also the magnetic field. We therefore expect the correlations between the angular fluctuations in the different radiation fields to provide  insights into the nature of turbulence and also the emission mechanism in each of the three bands considered here.  We have considered the  cross angular power spectra  $C^{TA}_\ell$,$C^{TB}_\ell$ and $C^{AB}_\ell$ to quantify the $\ell$ dependence of the correlations,  and the dimensionless correlation coefficients  $c^{TA}_\ell$,$c^{TB}_\ell$ and $c^{AB}_\ell$ to quantify the strength of the correlations. Considering $c^{TA}_\ell$, we see  (Figure $\ref{fig:corr_coeff}$) that this has very small positive  values $(\sim 10^{-4})$ which indicates that radio and X-ray A are extremely weakly correlated. Further,    $C^{TA}_\ell$ shows several breaks in the range $\ell^{TA}_m- \ell^{TA}_b$ and a single power law or a  power law with a single break does not provide a good fit to the data.  In contrast,  $c^{TB}_\ell$ and $c^{AB}_\ell$ both have  values in the range $0.4-0.8$ indicating strong correlations between radio and X-ray B, and X-ray A and B respectively. Considering   $C^{TB}_\ell$ we see that this may be  fitted with a broken power law. In the range $\ell < \ell^{TB}_b=1.16 \times 10^5 $, the power law slope $\beta=-2.11$ is close to the  power law $\beta=-2.28$ observed for radio, whereas at $\ell \ge \ell^{TB}_b$,  $C^{TB}_\ell$ follows $\beta=-2.76$ which roughly matches with the power law scaling $\beta=-2.62$ of $C^{BB}_\ell$. We may interpret this as indicating that the emission sources for X-ray B roughly follow the angular distribution of the radio sources at small $\ell$ or large angular scales whereas the radio sources roughly follow the angular distribution of the X-ray B sources at large $\ell$ or small angular scales.  Furthermore, $\ell^{TB}_b$ is approximately the geometric mean of $\ell^T_b$ and $\ell^B_b$.  Considering $C^{AB}_\ell$,  this has a power law slopes $\beta=-2.11$ for $\ell<\ell^{AB}=2.11\times10^5$ and $\beta=-2.70$ for $\ell \geq \ell^{AB}_b$. These 
$\beta$ values roughly match  the mean values of the slopes for X-ray A and X-ray B at low $\ell$ and high $\ell$ respectively. Moreover, $\ell^{AB}_b$ is roughly a little smaller   than the mean of $\ell^A_b$ and $\ell^B_b$. To summarize the results of the correlation analysis, we find that X-ray B is correlated with both radio and X-ray A while the latter two are not correlated among themselves.    It is well known that the radio emission is predominantly non-thermal synchrotron radiation  while X-ray A is believed to be mostly thermal bremsstrahlung radiation. Our study shows that the angular distribution of these two respective emission sources is largely uncorrelated.
This possibly originates from the fact that  the  synchrotron radiation comes from the  regions associated with the forward shock while  most of the thermal bremsstrahlung in X-ray A comes from ejecta shocked by the reverse shock. 
Interestingly, X-ray B is well correlated with both radio and X-ray A. This supports the possibility  that X-ray B is  a combination of thermal bremsstrahlung and non-thermal synchrotron radiation.

In conclusion we note that nearly all of the power spectra measured here, both auto and cross, can be modelled with broken power laws where the slope steepens at $\ell$ larger than the break. 
The various power spectra measured here provides a wealth of information regarding the nature of turbulence and the structure of the SNR.  For radio, we interpret this in terms of MHD  turbulence in the plasma of the SNR.  Our simplistic model indicates that we may possibly interpret the break in the radio power spectrum in terms of the  geometry of the SNR shell structure. 
A more detailed modeling of the SNR  taking into account inputs from all the various power spectra measured here is beyond the scope of the present work.  
This requires detailed simulations and theory  which will be carried out in the future.

  The theoretical prediction for 2D turbulence by \cite{Kraichnan1967, kraichnan_1971} gives an alternative description for the observed break of the various power spectra presented here. The change of the power law slope can be interpreted to occur due to energy transfer in an inverse cascade to larger length scales and a cascade of enstrophy to smaller length scales in response to the energy injection at the break. The ejecta and CSM interacting with the shocks possibly contributes to the input of energy and vorticity \citep{Beresnyak_2009} at a length scale corresponding to the shell thickness of Cas A SNR. The verification of the above conjectures needs detailed simulations which will be attempted in a separate work.

\section*{Acknowledgements}
This research has made use of data obtained from the Chandra Data Archive and the Chandra Source Catalog, and software provided by the Chandra X-ray Center (CXC) in the application packages CIAO, and Sherpa. NR thanks Harvard-Smithsonian Center for Astrophysics for hospitality during a visit and stay when part of this work was done. Support for HMG was provided by the National Aeronautics and Space Administration through the Smithsonian Astrophysical Observatory contract SV3-73016 to MIT for Support of the Chandra X-Ray Center, which is operated by the Smithsonian Astrophysical Observatory for and on behalf of the National Aeronautics Space Administration under contract NAS8-03060. We thank Martin Laming for a very useful review.

\section*{Data Availability}
The datasets used in this article were derived from sources in the public domain: [radio: NRAO Science Data Archive, \url{https://archive.nrao.edu/archive/advquery.jsp} and X-ray: Chandra Data Archive, \url{https://cxc.harvard.edu/cda/} ].
   
\bibliographystyle{mnras}

\bibliography{bibreportext}

\label{lastpage}

\end{document}